\documentclass[aps,onecolumn]{revtex4}
\usepackage{amssymb,epsf}
\usepackage{latexsym}

\begin{document}

\title{Phase transition of charged Black Holes in Brans-Dicke theory \\
through geometrical thermodynamics}
\author{S. H. Hendi$^{1,2}$\footnote{email address: hendi@shirazu.ac.ir},
S. Panahiyan$^{1,3}$\footnote{email address: sh.panahiyan@gmail.com}, B. Eslam Panah$^1$\footnote{%
email address: behzad.eslampanah@gmail.com} and Z. Armanfard$^1$\footnote{%
email address: zahra.armanfard@gmail.com}}
\affiliation{$^1$Physics Department and Biruni Observatory,
College of Sciences, Shiraz
University, Shiraz 71454, Iran\\
$^2$Research Institute for Astronomy and Astrophysics of Maragha
(RIAAM),
Maragha, Iran\\
$^3$ Physics Department, Shahid Beheshti University, Tehran 19839,
Iran}

\begin{abstract}
In this paper, we take into account black hole solutions of
Brans-Dicke-Maxwell theory and investigate their stability and
phase transition points. We apply the concept of geometry in
thermodynamics to obtain phase transition points and compare its
results with those, calculated in canonical ensemble through heat
capacity. We show that these black holes enjoy second order phase
transitions. We also show that there is a lower bound for the
horizon radius of physical charged black holes in Brans-Dicke
theory which is originated from restrictions of positivity of
temperature. In addition, we find that employing specific
thermodynamical metric in the context of geometrical
thermodynamics, yields divergencies for thermodynamical Ricci
scalar in places of the phase transitions. It will be pointed out
that due to characteristics behavior of thermodynamical Ricci
scalar around its divergence points, one is able to distinguish
the physical limitation point from the phase transitions. In
addition, the free energy of these black holes will be obtained
and its behavior will be investigated. It will be shown that the
behavior of the free energy in the place where the heat capacity
diverges, demonstrates second order phase transition
characteristics.
\end{abstract}

\maketitle

\section{Introduction\label{Intro}}

Einstein's general relativity is able to describe the dynamics of
our solar system well enough. In addition, this theory predicted
the existence of gravitational waves that recently, was observed
by LIGO and Virgo collaboration \cite{LIGO}. It is notable that,
observations of the LIGO collaboration could be employed to test
the validation of Einstein theory of the gravity or the necessity
of modified theories of gravity \cite{Corda}. Einstein theory of
gravity may have some problems in describing gravity accurately at
all scales. One of the problems that general relativity faced was
that, it couldn't describe the accelerated expansion of the
universe accurately \cite{Perlmutter}. Also, it is consistent with
neither Mach's principle nor Dirac's large number hypothesis
\cite{DLNH}. Herein,
cosmologists explored various alternatives for gravitational fields \cite%
{Clifton,FR}. The pioneers of studying on scalar-tensor theory were done by
Brans and Dicke \cite{Brans}. This theory accommodates both Mach's principle
and Dirac's large number hypothesis \cite{DLNH}. Scientists have
investigated various aspects of black holes and gravitational collapse in
Brans-Dicke (BD) theory due to their importance in both classical and
quantum aspects of gravity \cite{Scheel}. It has been shown that the
stationary and vacuum BD solution in $4-$dimensions is just the Kerr
solution with a trivial scalar field everywhere \cite{Hawking}. Cai and
Myung showed that in $4-$dimensions the BD-Maxwell (BDM) solution is just
the Reissner-Nordstr\"{o}m (RN) solution with a constant scalar field \cite%
{Cai}. However in higher dimensions, due to the fact that the action of
Maxwell field is not invariant under conformal transformation and the stress
energy tensor of Maxwell field is not traceless, the solution of BDM is RN
solution with a nontrivial scalar field.

On the other hand, the black hole thermodynamics has been the field of
interest for many researchers since the work of Hawking \cite{Hawking1975}.
Recently, in most treatments of black hole thermodynamics, physicists have
considered the cosmological constant $(\Lambda)$ as a dynamical variable
\cite{Henneaux,Teitelboim}. Furthermore, some authors suggested to treat $%
\Lambda $ as a thermodynamic variable \cite{Sekiwa}, such as thermodynamical
pressure \cite{Brown,Dolan,PVReferences,HendiArman}. One of the interesting
aspects of the black hole thermodynamics is their stability. In order to
have a black hole in thermal stability, its heat capacity must be positive.
In other words, positivity of the heat capacity guarantees thermal stability
of the black holes. This approach of studying the stability is in the
context of the canonical ensemble. Studying the heat capacity of a system
also provides a mechanism to study the phase transitions of that system.
There are two distinctive points: in one point, changing in the sign of the
heat capacity is denoted as a physical limitation point and therefore, the
root of the heat capacity is a border between non-physical and physical
black hole solutions. Other characterized points are related to divergencies
of the heat capacity. We identify these points as second order phase
transitions \cite{Davies}.

In the past few decades, applying the thermodynamical geometry for studying
the phase transition of black holes has gained lots of attentions. These
studies were pioneered by Weinhold \cite{Weinhold} and Ruppeiner \cite%
{Ruppeiner}. Weinhold introduced a metric on the space of equilibrium state
and defined the metric tensor as the second derivatives of internal energy
with respect to entropy and other extensive quantities. On the other hand,
the metric that Ruppeiner introduced was defined as the minus second
derivatives of entropy with respect to internal energy and other extensive
quantities, which was conformal to Weinhold's metric \cite{Salamon}.
However, applying these treatments to study black hole thermodynamics caused
some puzzling anomalies. Neither Weinhold nor Ruppeiner metrics were
invariant under Legendre transformation. A few years ago, Quevedo \cite%
{Quevedo} proposed an approach to obtain a metric which was
Legendre invariant in the space of equilibrium state. His work was
based on the observation in which standard thermodynamics was
invariant with respect to Legendre transformation. The formalism
of geometrothermodynamics (GTD) indicates that phase transition
occurs at points where the thermodynamics curvature is singular,
as a consequence, the curvature can be interpreted as a measure of
thermodynamics interaction. After Quevedo, interesting studies
were followed by some authors \cite{CaiC}. Recently, it was
pointed out that using the mentioned approaches toward GTD may
sometimes confront with some problems. In order to overcome these
problems, a new approach was introduced in Ref. \cite{NewMetric}.

The outline of the paper will be as follow; in Sec. \ref{BDM}, we review the
charged BD black holes and their thermodynamical quantities. In Sec. \ref%
{Thermo}, we introduce the approaches for studying phase transitions of
these black holes in the context of heat capacity and geometrical
thermodynamics. Then, we investigate the existence of the phase transitions
in the context of two mentioned approaches and compare them with each other.
We also investigate the effect of BD parameter. In addition, we study the
free energy of BD black holes as well. The last section is devoted to
closing remarks.

\section{Black holes solutions in BDM gravity \label{BDM}}

Regarding $(n+1)$-dimensional BDM theory, one finds related action as \cite%
{Cai}
\begin{equation}
I_{G}=-\frac{1}{16\pi }\int_{\mathcal{M}}d^{n+1}x\sqrt{-g}\left( \Phi
\mathcal{R}-\frac{\omega }{\Phi }(\nabla \Phi )^{2}-V(\Phi )-F_{\mu \nu
}F^{\mu \nu }\right) ,  \label{acBD}
\end{equation}%
where $\Phi $ and $V(\Phi )$ are, respectively, a scalar field and its
self-interacting potential. Besides, the factor $\omega $ is the coupling
constant, $\mathcal{R}$ is the scalar curvature, $F_{\mu \nu }=\partial
_{\mu }A_{\nu }-\partial _{\nu }A_{\mu }$ is the electromagnetic tensor
field and $A_{\mu }$ is the electromagnetic potential. Equations of motion
can be obtained with the following explicit forms by varying the action (\ref%
{acBD}) with respect to the gravitational field $g_{\mu \nu }$, the scalar
field $\Phi $ and the gauge field $A_{\mu }$ \cite{Cai}
\begin{eqnarray}
G_{\mu \nu } &=&\frac{\omega }{\Phi ^{2}}\left( \nabla _{\mu }\Phi \nabla
_{\nu }\Phi -\frac{1}{2}g_{\mu \nu }(\nabla \Phi )^{2}\right) -\frac{V(\Phi )%
}{2\Phi }g_{\mu \nu }+\frac{1}{\Phi }\left( \nabla _{\mu }\nabla _{\nu }\Phi
-g_{\mu \nu }\nabla ^{2}\Phi \right)  \nonumber \\
&&+\frac{2}{\Phi }\left( F_{\mu \lambda }F_{\nu }^{\lambda }-\frac{1}{4}%
F_{\rho \sigma }F^{\rho \sigma }g_{\mu \nu }\right) ,  \label{field01}
\end{eqnarray}%
\begin{equation}
\nabla ^{2}\Phi =-\frac{n-3}{2\left[ \left( n-1\right) \omega +n\right] }%
F^{2}+\frac{1}{2\left[ \left( n-1\right) \omega +n)\right] }\left[ (n-1)\Phi
\frac{dV(\Phi )}{d\Phi }-\left( n+1\right) V(\Phi )\right] ,  \label{field02}
\end{equation}%
\begin{equation}
\nabla _{\mu }F^{\mu \nu }=0,  \label{field03}
\end{equation}%
where $G_{\mu \nu }$ and $\nabla _{\mu }$ are, respectively, the Einstein
tensor and covariant derivative of manifold $\mathcal{M}$ with metric $%
g_{\mu \nu }$. Due to the appearance of inverse powers of the scalar field
on the right hand side of (\ref{field01}), solving the field equations (\ref%
{field01})-(\ref{field03}), directly, is a non-trivial task. This difficulty
can be removed, by using a suitable conformal transformation \cite{Cai}.
Indeed, via the conformal transformation the BDM theory can be transformed
into the Einstein-Maxwell theory with a minimally coupled scalar dilaton
field. Suitable conformal transformation can be shown as \cite{Cai}
\begin{eqnarray}
\bar{g}_{\mu \nu } &=&\Phi ^{2/(n-1)}g_{\mu \nu },   \\
\bar{\Phi} &=&\frac{n-3}{4\alpha }\ln \Phi ,  \label{con}
\end{eqnarray}%
where
\begin{equation}
\alpha =(n-3)/\sqrt{4(n-1)\omega +4n}.  \label{a}
\end{equation}

It is worth mentioning that all functions and quantities in Jordan frame (${g%
}_{\mu \nu }$, ${\Phi }$ and ${F}_{\mu \nu }$) can be transformed into
Einstein frame ($\bar{g}_{\mu \nu }$, $\bar{\Phi}$ and $\bar{F}_{\mu \nu }$%
). Applying the mentioned conformal transformation on the BD action (\ref%
{acBD}), one finds the action of dilaton gravity
\begin{equation}
\bar{I}_{G}=-\frac{1}{16\pi }\int_{\mathcal{M}}d^{n+1}x\sqrt{-\bar{g}}%
\left\{ \bar{\mathcal{R}}-\frac{4}{n-1}(\bar{\nabla}\bar{\Phi})^{2}-\bar{V}(%
\bar{\Phi})-\exp \left( -\frac{4\alpha \bar{\Phi}}{(n-1)}\right) \bar{F}%
_{\mu \nu }\bar{F}^{\mu \nu }\right\} ,  \label{con-ac}
\end{equation}%
where $\bar{\nabla}$ and $\overline{\mathcal{R}}$ are, respectively, the
covariant derivative and Ricci scalar corresponding to the metric $\bar{g}%
_{\mu \nu }$, and $\bar{V}(\bar{\Phi})$ is
\begin{equation}
\bar{V}(\bar{\Phi})=\Phi ^{-(n+1)/(n-1)}V(\Phi ).  \label{poten}
\end{equation}

Regarding $(n+1)$-dimensional Einstein-Maxwell-dilaton action (\ref{con-ac}%
), $\alpha $ is an arbitrary constant that governs the strength of coupling
between the dilaton and Maxwell fields. One can obtain the equations of
motion by varying this action (\ref{con-ac}) with respect to $\bar{g}_{\mu
\nu }$, $\bar{\Phi}$ and $\bar{F}_{\mu \nu }$
\begin{equation}
\bar{\mathcal{R}}_{\mu \nu }=\frac{4}{n-1}\left( \bar{\nabla}_{\mu }\bar{\Phi%
}\bar{\nabla}_{\nu }\bar{\Phi}+\frac{1}{4}\bar{V}\bar{g}_{\mu \nu }\right)
+2e^{-4\alpha \bar{\Phi}/(n-1)}\left( \bar{F}_{\mu \lambda }\bar{F}_{\nu
}^{\lambda }-\frac{1}{2(n-1)}\bar{F}_{\rho \sigma }\bar{F}^{\rho \sigma }%
\bar{g}_{\mu \nu }\right) ,  \label{fieldc1}
\end{equation}%
\begin{equation}
\bar{\nabla}^{2}\bar{\Phi}=\frac{n-1}{8}\frac{\partial \bar{V}}{\partial
\bar{\Phi}}-\frac{\alpha }{2}e^{-4\alpha \bar{\Phi}/(n-1)}\bar{F}_{\rho
\sigma }\bar{F}^{\rho \sigma },  \label{fieldc2}
\end{equation}%
\begin{equation}
\partial _{\mu }\left[ \sqrt{-\bar{g}}e^{-4\alpha \bar{\Phi}/(n-1)}\bar{F}%
^{\mu \nu }\right] =0.  \label{fieldc3}
\end{equation}

Assuming the $\left( \bar{g}_{\mu \nu },\bar{F}_{\mu \nu },\bar{\Phi}\right)
$ as solutions of Eqs. (\ref{fieldc1})-(\ref{fieldc3}) with potential $\bar{V%
}\left( \bar{\Phi}\right) $ and comparing Eqs. (\ref{field01})-(\ref{field03}%
) with Eqs. (\ref{fieldc1})-(\ref{fieldc3}), the solutions of Eqs. (\ref%
{field01})-(\ref{field03}) with potential $V(\Phi )$ can be written as
\begin{equation}
\left[ g_{\mu \nu },F_{\mu \nu },\Phi \right] =\left[ \exp \left( -\frac{%
8\alpha \bar{\Phi}}{\left( n-1\right) (n-3)}\right) \bar{g}_{\mu \nu },\bar{F%
}_{\mu \nu },\exp \left( \frac{4\alpha \bar{\Phi}}{n-3}\right) \right] .
\label{sol}
\end{equation}

As a consequence, we can solve Eqs. (\ref{fieldc1})-(\ref{fieldc3}) with
suitable potential, instead of solving Eqs. (\ref{field01})-(\ref{field03}).
Assuming an $(n+1)$-dimensional static and spherically symmetric metric
\begin{equation}
d\bar{s}^{2}=-f(r)dt^{2}+\frac{dr^{2}}{f(r)}+r^{2}R^{2}(r)d\Omega _{n-1}^{2},
\label{metric}
\end{equation}%
where $d\Omega _{n-1}^{2}$ is the metric of a unit $(n-1)$-sphere, and $f(r)$
and $R(r)$ are metric functions. By integrating the Maxwell equation (\ref%
{fieldc3}), we can obtain the nonzero electric field $\bar{F}_{tr}$ as
\begin{equation}
\bar{F}_{tr}=\frac{q}{(rR)^{n-1}}\exp \left( \frac{4\alpha \bar{\Phi}}{n-1}%
\right) ,  \label{maxwell}
\end{equation}%
where $q$ is an integration constant related to electric charge.
Now, we regard the following Liouville-type potential to solve the field
equations
\begin{equation}
\bar{V}(\bar{\Phi})=2\Lambda \exp \left( \frac{4\alpha \bar{\Phi}}{n-1}%
\right) +\frac{(n-1)(n-2)\alpha ^{2}}{c^{2}\left( \alpha ^{2}-1\right) }e^{%
\frac{4\bar{\Phi}}{(n-1)\alpha }}.  \label{liovilpoten}
\end{equation}


Taking into account the metric (\ref{metric}) with Maxwell field (\ref%
{maxwell}) and potential (\ref{liovilpoten}), the consistent solutions of
Eqs. (\ref{fieldc1}) and (\ref{fieldc2}) are \cite{HendiArman}
\begin{eqnarray}
f(r) &=&-\frac{\left( n-2\right) \left( \alpha ^{2}+1\right) ^{2}c^{-2\gamma
}r^{2\gamma }}{\left( \alpha ^{2}+n-2\right) \left( \alpha ^{2}-1\right) }+%
\frac{2\Lambda (\alpha ^{2}+1)^{2}c^{2\gamma }}{(n-1)(\alpha ^{2}-n)}%
r^{2(1-\gamma )}-\frac{m}{r^{(n-2)}}r^{(n-1)\gamma }  \nonumber \\
&&+\frac{2q^{2}(\alpha ^{2}+1)^{2}c^{-2(n-2)\gamma }}{(n-1)(\alpha
^{2}+n-2)r^{2(n-2)(1-\gamma )}},  \label{F(r)} \\
R(r) &=&\exp (\frac{2\alpha \bar{\Phi}}{n-1})=\left( \frac{c}{r}\right)
^{\gamma },  \label{R(r)} \\
\bar{\Phi}(r) &=&\frac{(n-1)\alpha }{2(1+\alpha ^{2})}\ln (\frac{c}{r}),
\label{phi}
\end{eqnarray}
where $m$ is an integration constant which is related to the total mass, $c$
is another arbitrary constant related to the scalar field and $\gamma=\alpha
^{2}/(1+\alpha ^{2})$.

Now, using the conformal transformation, we are able to obtain the solutions
of Eqs. (\ref{field01})--(\ref{field03}). Considering the following
spherically symmetric metric
\begin{equation}
ds^{2}=-U(r)dt^{2}+\frac{dr^{2}}{V(r)}+r^{2}H^{2}(r)d\Omega _{n-1}^{2},
\label{metric1}
\end{equation}%
with Eqs. (\ref{field01})-(\ref{field03}), we find that the functions $U(r)$
and $V(r)$ are \cite{HendiArman}
\begin{eqnarray}
U(r) &=&\frac{2\Lambda (\alpha ^{2}+1)^{2}c^{2\gamma (\frac{n-5}{n-3})}}{%
(n-1)(\alpha ^{2}-n)}r^{2(1-\frac{\gamma \left( n-5\right) }{n-3})}-\frac{%
mc^{(\frac{-4\gamma }{n-3})}}{r^{(n-2)}}r^{\gamma (n-1+\frac{4}{n-3})}
\nonumber \\
&&+\frac{2q^{2}(\alpha ^{2}+1)^{2}c^{-2\gamma (n-2+\frac{2}{n-3})}}{%
(n-1)(\alpha ^{2}+n-2)r^{2[(n-2)(1-\gamma )-\frac{2\gamma }{n-3}]}}-\frac{%
\left( n-2\right) \left( \alpha ^{2}+1\right) ^{2}}{\left( \alpha
^{2}+n-2\right) \left( \alpha ^{2}-1\right) }\left( \frac{c}{r}\right)
^{-2\gamma \left( \frac{n-1}{n-3}\right) },  \label{u(r)} \\
V(r) &=&\frac{2\Lambda (\alpha ^{2}+1)^{2}c^{2\gamma (\frac{n-1}{n-3})}}{%
(n-1)(\alpha ^{2}-n)}r^{2(1-\frac{\gamma \left( n-1\right) }{n-3})}-\frac{%
mc^{(\frac{4\gamma }{n-3})}}{r^{(n-2)}}r^{\gamma (n-1-\frac{4}{n-3})}
\nonumber \\
&&+\frac{2q^{2}(\alpha ^{2}+1)^{2}c^{-2\gamma (n-2-\frac{2}{n-3})}}{%
(n-1)(\alpha ^{2}+n-2)r^{2[(n-2)(1-\gamma )+\frac{2\gamma }{n-3}]}}-\frac{%
\left( n-2\right) \left( \alpha ^{2}+1\right) ^{2}}{\left( \alpha
^{2}+n-2\right) \left( \alpha ^{2}-1\right) }\left(
\frac{c}{r}\right) ^{-2\gamma \left( \frac{n-5}{n-3}\right) },
\label{V(r)}
\end{eqnarray}%
where we used the conformal transformation of Eq. (\ref{liovilpoten}) with
the following explicit form
\begin{equation}
V(\Phi )=2\Lambda \Phi ^{2}+\frac{(n-1)(n-2)\alpha
^{2}}{c^{2}\left( \alpha ^{2}-1\right) }\Phi ^{\lbrack
(n+1)(1+\alpha ^{2})-4]/[(n-1)\alpha ^{2}]}.
\end{equation}

In addition, one can use the conformal transformation to obtain consistent
electromagnetic field as
\begin{equation}
F_{tr}=\frac{qc^{(3-n)\gamma }}{r^{(n-3)(1-\gamma )+2}}.  \label{ftr}
\end{equation}

As one can see, the electromagnetic field becomes zero as $r\longrightarrow
\infty $. It is evident that as $\omega \longrightarrow \infty$, the
obtained solutions are just the charged solutions of Einstein gravity (RN
AdS black hole).

Using the Euclidian action, the finite mass and the entropy of the black
hole can be obtained \cite{Cai}
\begin{eqnarray}
M &=&\frac{c^{(n-1)\gamma }}{16\pi }\left( \frac{n-1}{1+\alpha ^{2}}\right)
m,  \label{m} \\
S &=&\frac{c^{(n-1)\gamma }}{4}r_{+}^{(n-1)\left( 1-\gamma \right) }.
\label{s}
\end{eqnarray}

By considering the flux of electric field at infinity, one can find the
total charge of this configuration as
\begin{equation}
Q=\frac{q}{4\pi }.  \label{Q}
\end{equation}

Calculations show that the Hawking temperature of a BD black hole on the
outer horizon $r_{+}$ is
\begin{equation}
T=\frac{\kappa }{2\pi }=\left. \frac{1}{4\pi }\sqrt{\frac{V}{U}}\left( \frac{%
dU}{dr}\right) \right\vert _{r=r_{+}},  \label{tp}
\end{equation}%
where $\kappa $ is the surface gravity. After some simplifications, we
obtain \cite{HendiArman}
\begin{eqnarray}
T &=&-\frac{2(1+\alpha ^{2})}{4\pi (n-1)}\left( \Lambda c^{2\gamma
}r_{+}^{1-2\gamma }+\frac{q^{2}c^{-2(n-2)\gamma }}{{r_{+}}^{\gamma }}{r_{+}}%
^{(2n-3)(\gamma -1)}\right)  \nonumber \\
&&+\frac{\left[ \gamma \left( n-3\right) -n+2\right] \left(
1-n\right) \left( n-2\right) }{2 r_{+}\left( \alpha
^{2}+n-2\right) \left( \alpha ^{2}-1\right) }\left(
\frac{c}{r_{+}}\right) ^{-2\gamma }.  \label{temp1}
\end{eqnarray}

\section{Stability, phase transition and geometrical thermodynamics \label%
{Thermo}}

In this section, first, we study stability and phase transition of
the solutions in the context of the heat capacity. Next, we
consider the geometrical approach for studying phase transition.
We investigate the effect of BD parameter and compare the results
of both approaches.

There are several approaches for studying the stability of black holes. One
of these approaches is related to studying the perturbed black holes and see
if and how they acquire stable state and will be in equilibrium. This
approach is known as dynamical stability of black holes. In this paper, we
are not interested in dynamical stability of black holes. We focus our
studies on the thermal stability of charged black hole solutions in the
context of BD theory through canonical ensemble. To do so, we calculate the
heat capacity and study its behavior.

Black holes should have a positive heat capacity in order to be thermally
stable. In other words, the positivity of the heat capacity guarantees the
local thermal stability of the black holes. One can use the following
relation for the heat capacity
\begin{equation}
C_{Q}=T\left( \frac{\partial ^{2}M}{\partial S^{2}}\right) _{Q}^{-1}=T\left(
\frac{\partial S}{\partial T}\right) _{Q}=T\left( \frac{\partial S}{\partial
r_{+}}\right) _{Q}\left( \frac{\partial T}{\partial r_{+}}\right) _{Q}^{-1}.
\label{CQ}
\end{equation}

On the other hand, it is possible to employ the heat capacity for
studying the phase transitions of black holes. In the context of
black holes, it is argued that root of the heat capacity
($C_{Q}=T=0$) is representing a border line between physical
($T>0$) and non-physical ($T<0$) black holes. We call it physical
limitation point. The system in case of this physical limitation
point has a change in sign of the heat capacity. In addition, it
is believed that the divergencies of the heat capacity are
representing phase transitions of black holes. These phase
transitions known as second order phase transition \cite{Davies}.
Therefore, the phase transition and limitation points of the black
holes in the context of the heat capacity are calculated with the
following relations
\begin{equation}
\left\{
\begin{array}{cc}
T=\left( \frac{\partial M}{\partial S}\right) _{Q}=0, & physical\text{ }%
limitation\text{ }point \\
&  \\
\left( \frac{\partial ^{2}M}{\partial S^{2}}\right) _{Q}=0, & second\text{ }%
order\ phase\text{ }transition%
\end{array}%
\right. .  \label{phase}
\end{equation}

Regarding Eq. (\ref{phase}) and in order to find physical limitation point,
one should solve the following equation for the entropy
\begin{eqnarray}
\left( \frac{\partial M}{\partial S}\right) _{Q} &=&7\frac{\pi ^{2}\Lambda
\left( n-9-4\omega \right) A_{1}c^{\left( n+1\right) j_{1}}}{4}+\frac{\left(
n-2\right) \pi ^{2}\left[ \left( n-1\right) \omega +n\right] A_{2}c^{\left(
n-3\right) j_{1}}}{2}  \nonumber \\
&&+\frac{\left( n-9-4\omega \right) Q^{2}}{c^{\left( n-3\right) j_{1}}A_{2}}%
=0,  \label{root}
\end{eqnarray}%
while as for the second order phase transition points (divergence points of
the heat capacity), we obtain the following relation%
\begin{eqnarray}
\left( \frac{\partial T}{\partial S}\right) _{Q}&=&\frac{\pi ^{2}\Lambda
\left( n-9-4\omega \right) A_{1}c^{\left( n+1\right) j_{1}}}{32}-\frac{%
\left( n-2\right) \pi ^{2}\left[ \left( n-1\right) \omega +n\right]
A_{2}c^{\left( n-3\right) j_{1}}}{16}  \nonumber \\
&&+\frac{\left[ \left( n-12\right) \omega +9\left( n-1\right) \right] Q^{2}}{%
8A_{2}c^{\left( n-3\right) j_{1}}}=0,  \label{dive}
\end{eqnarray}%
where
\begin{eqnarray}
A_{1}&=&\left( c^{-\frac{\left( n-3\right) ^{2}}{\left( 4n-4\right) \omega
+4n}}\left( 4S\right) ^{\frac{n^{2}+\left( 4\omega -2\right) n+9-4\omega }{4%
\left[ \left( \omega +1\right) n-\omega \right] \left( n-1\right) }}\right)
^{j_{2}},  \nonumber
\end{eqnarray}%
\begin{eqnarray}
A_{2}&=&\left( c^{-\frac{\left( n-3\right) ^{2}}{\left( 4n-4\right) \omega
+4n}}\left( 4S\right) ^{\frac{n^{2}+\left( 4\omega -2\right) n+9-4\omega }{4%
\left[ \left( \omega +1\right) n-\omega \right] \left( n-1\right) }}\right)
^{j_{3}},  \nonumber
\end{eqnarray}%
\begin{eqnarray}
j_{1}&=&\zeta \left( n-3\right) ^{2},  \nonumber
\end{eqnarray}%
\begin{eqnarray}
j_{2}&=&\zeta \left[ 4n\omega \left( n-1\right) +3n\left( n+1\right) +3n-9%
\right],  \nonumber
\end{eqnarray}%
\begin{eqnarray}
j_{3}&=&\zeta \left( n-1\right) \left[ \left( 4n-8\right) \omega +5n-9\right]
,  \nonumber
\end{eqnarray}%
\begin{eqnarray}
\zeta &=&\left[ \left( n-1\right) ^{2}+4\left( n-1\right) \omega +8\right]
^{-1}.  \nonumber
\end{eqnarray}

\begin{figure}[tbp]
$%
\begin{array}{cc}
\epsfxsize=6cm \epsffile{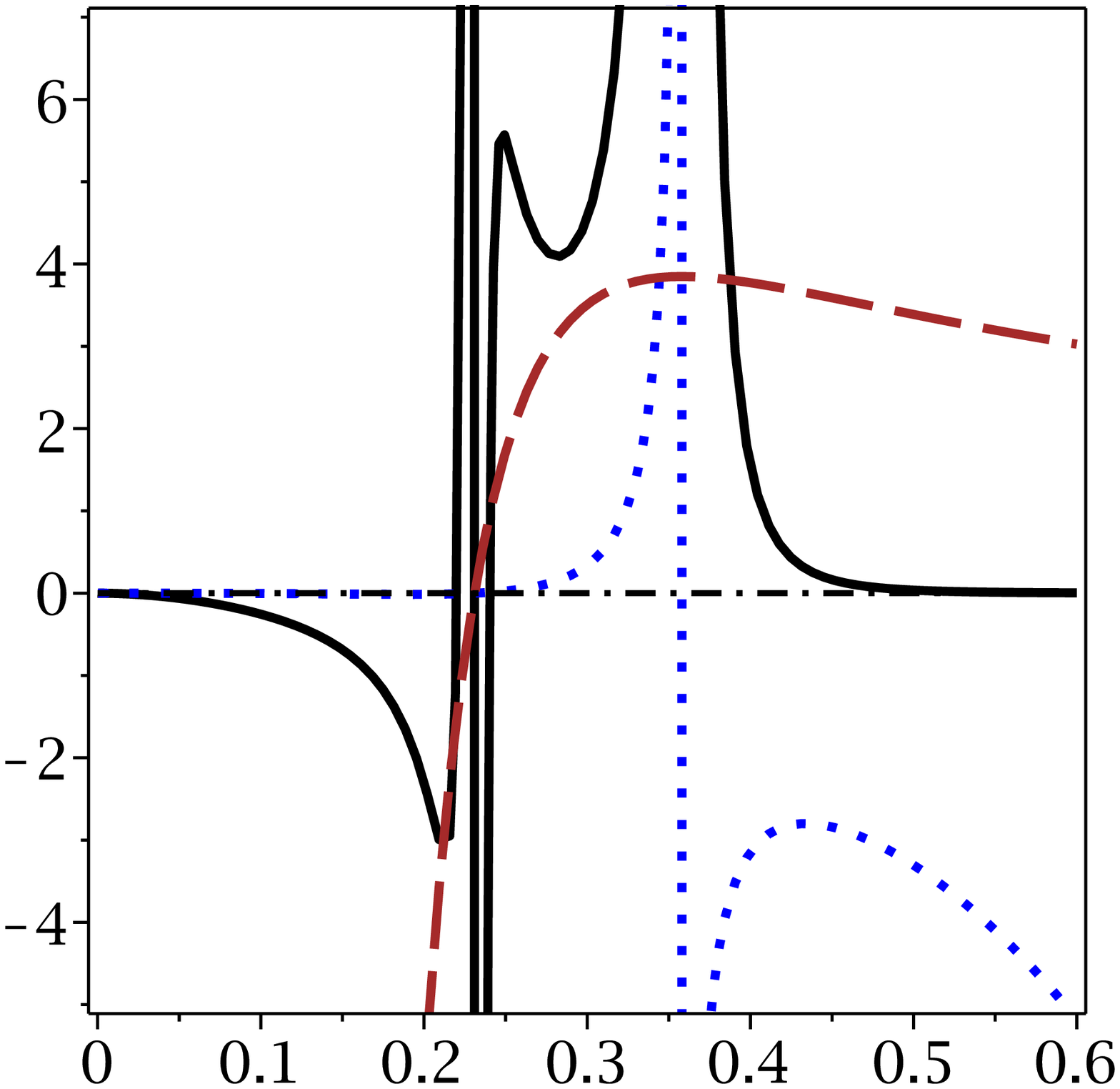} & \epsfxsize=6cm \epsffile{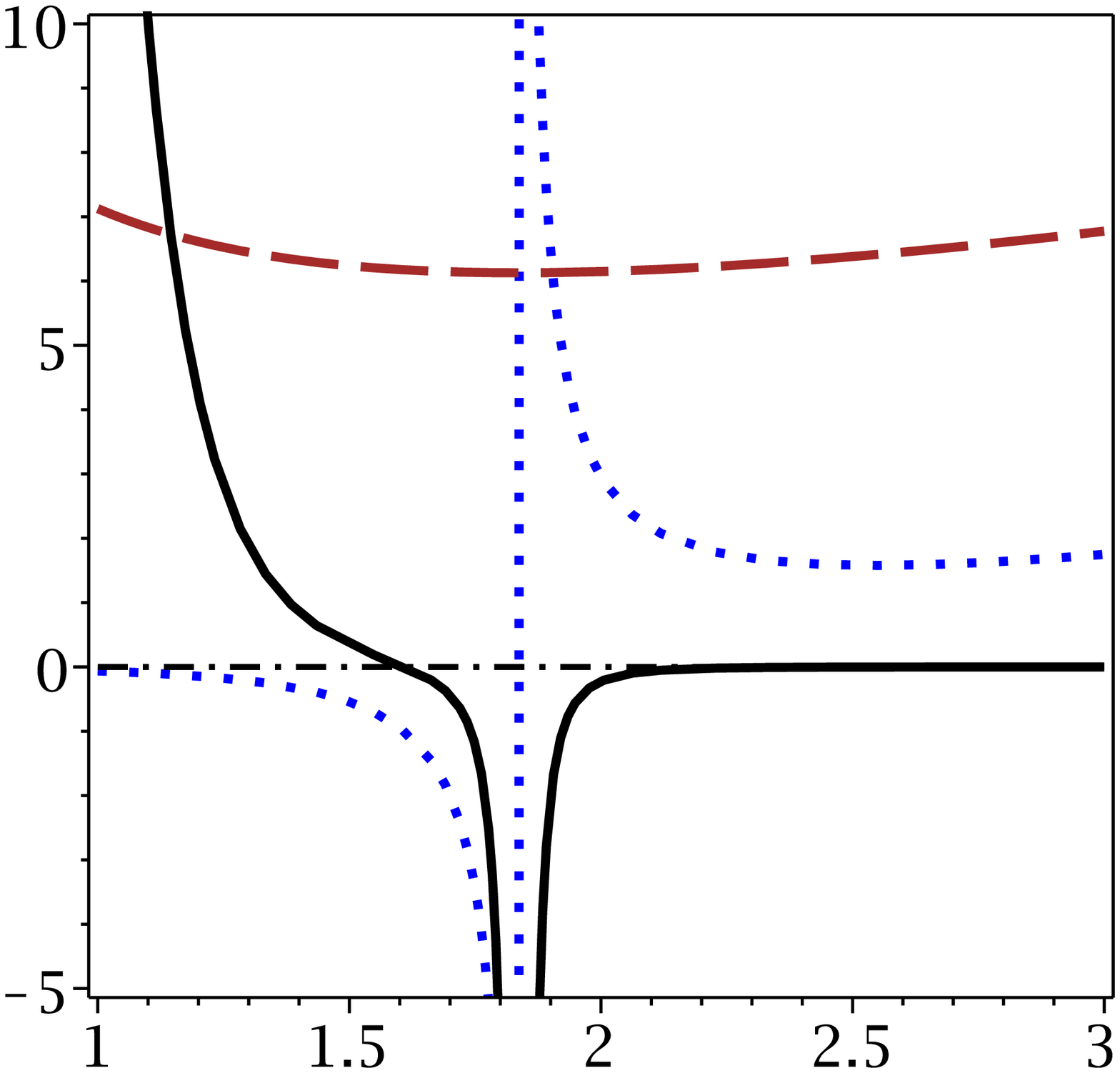}
\\
\epsfxsize=6cm \epsffile{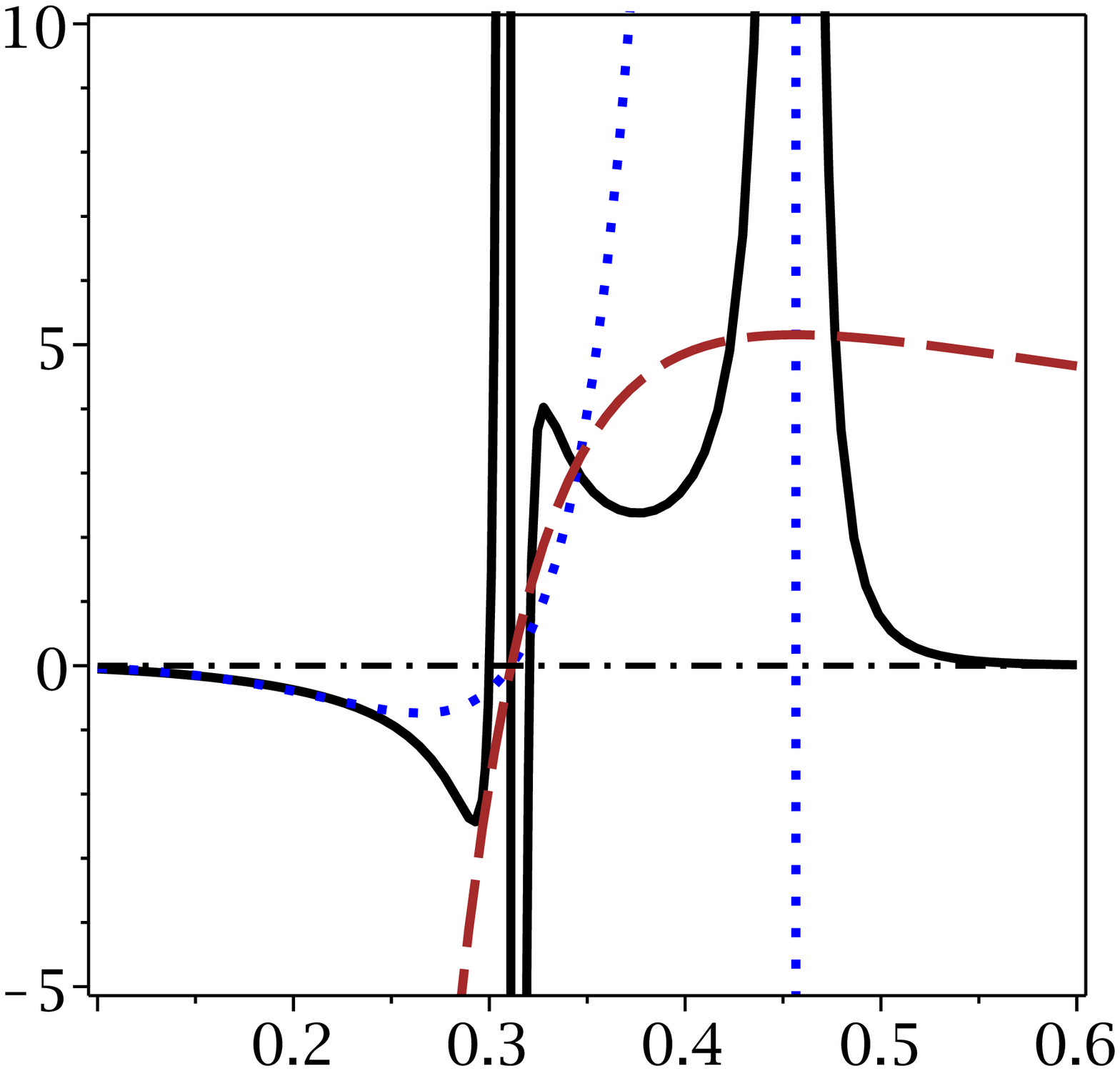} & \epsfxsize=6cm \epsffile{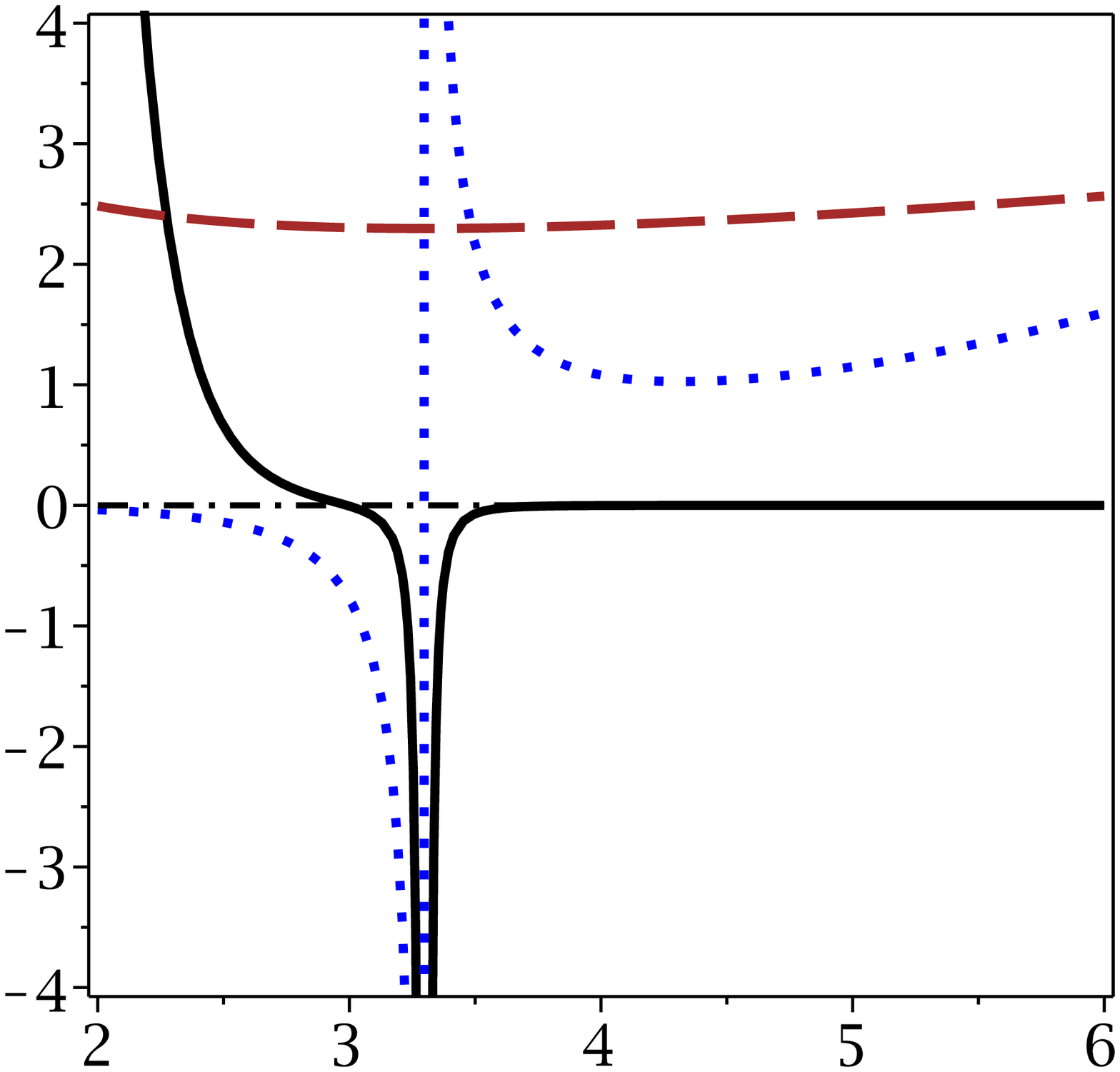}%
\end{array}
$%
\caption{For different scales: $\mathcal{R}$ (continuous line), $C_{Q}$
(dotted line) and $T$ (dashed line) versus $r_{+}$ for $\Lambda =-1$, $%
\protect\omega=1$, $c=1$ and $q=0.1$. (up diagrams: $d=5$ \& down
diagrams: $d=6$) } \label{Fig1}
\end{figure}

\begin{figure}[tbp]
$%
\begin{array}{ccc}
\epsfxsize=5cm \epsffile{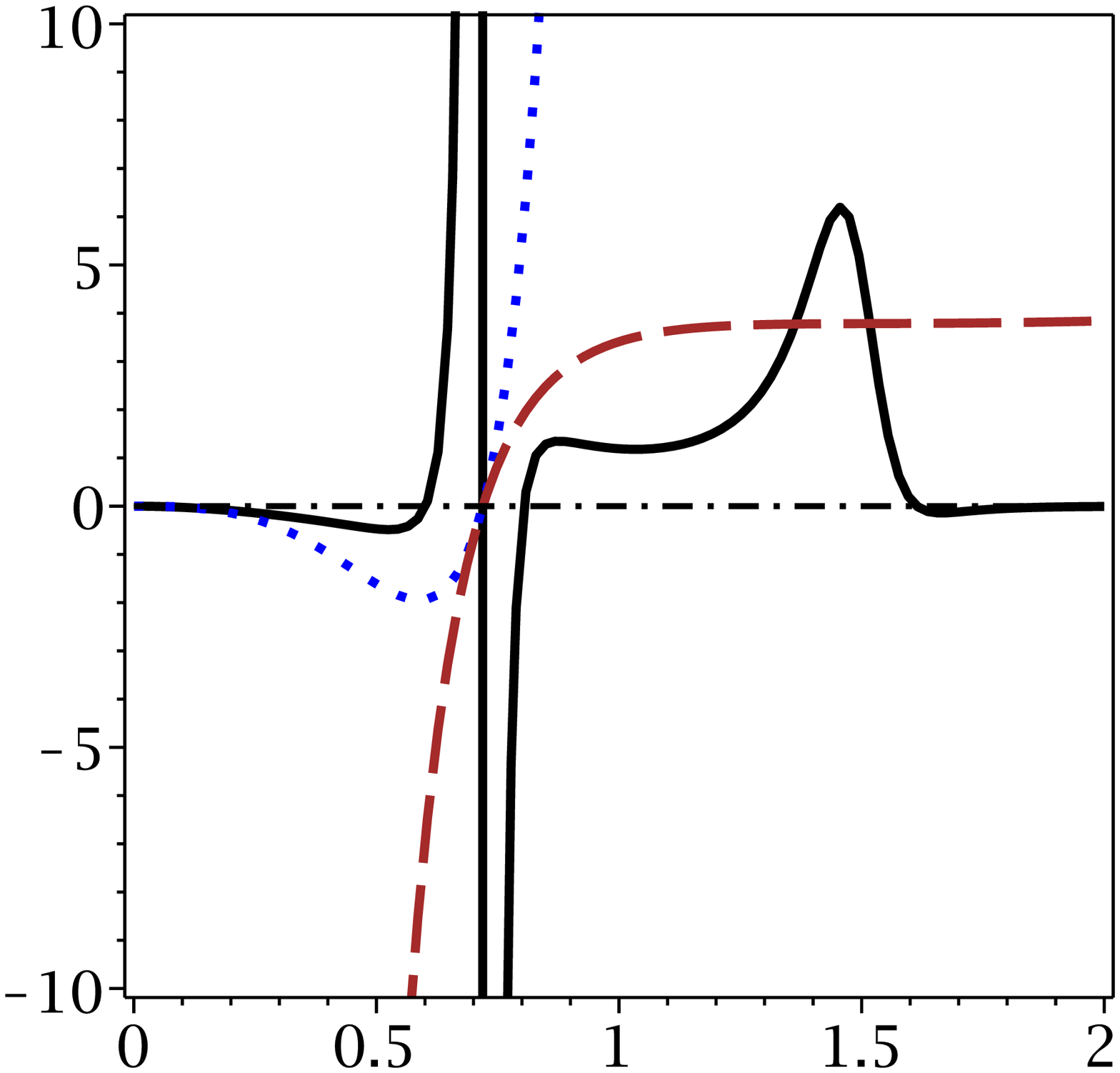} & \epsfxsize=5cm \epsffile{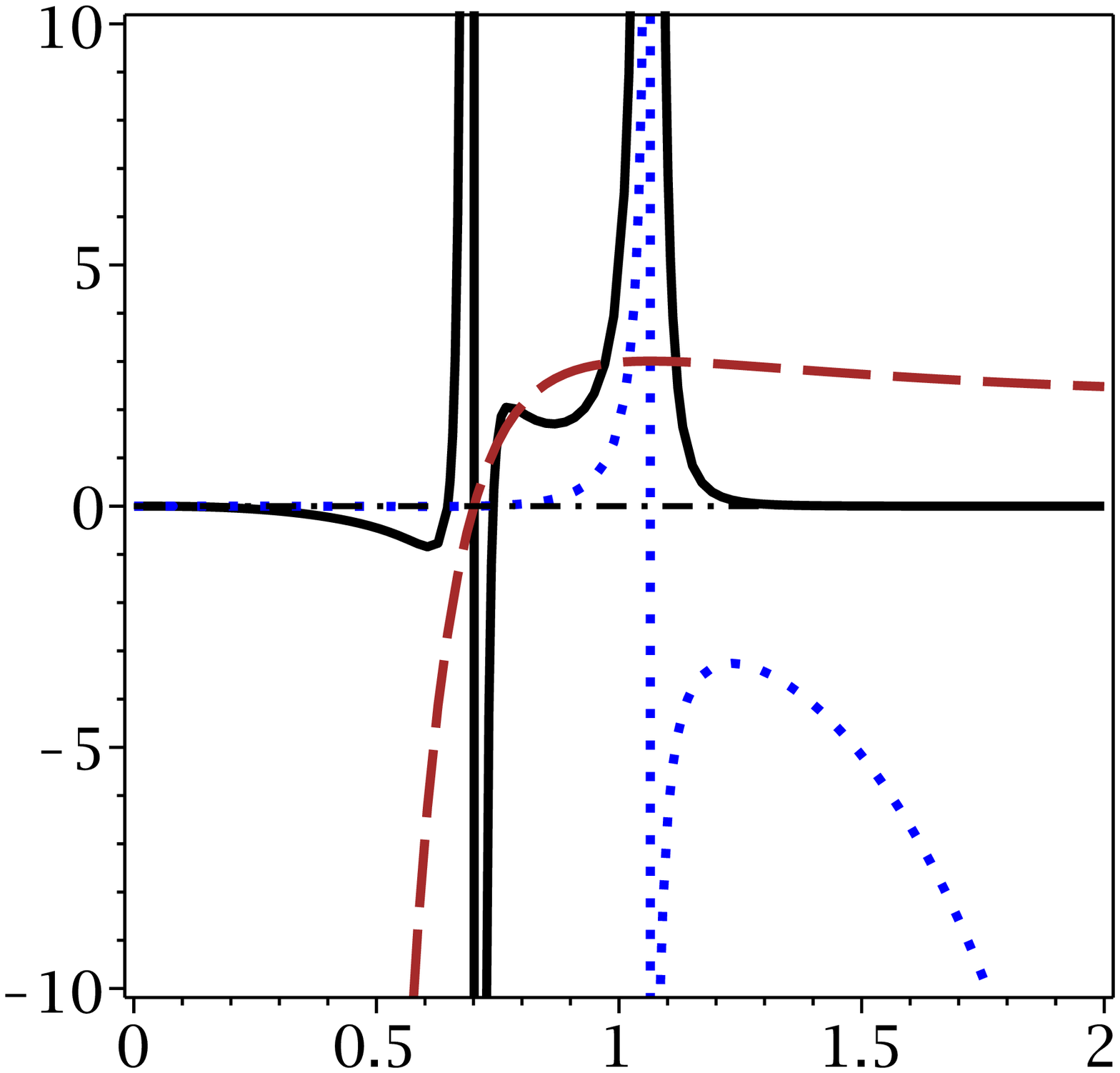} & %
\epsfxsize=5cm \epsffile{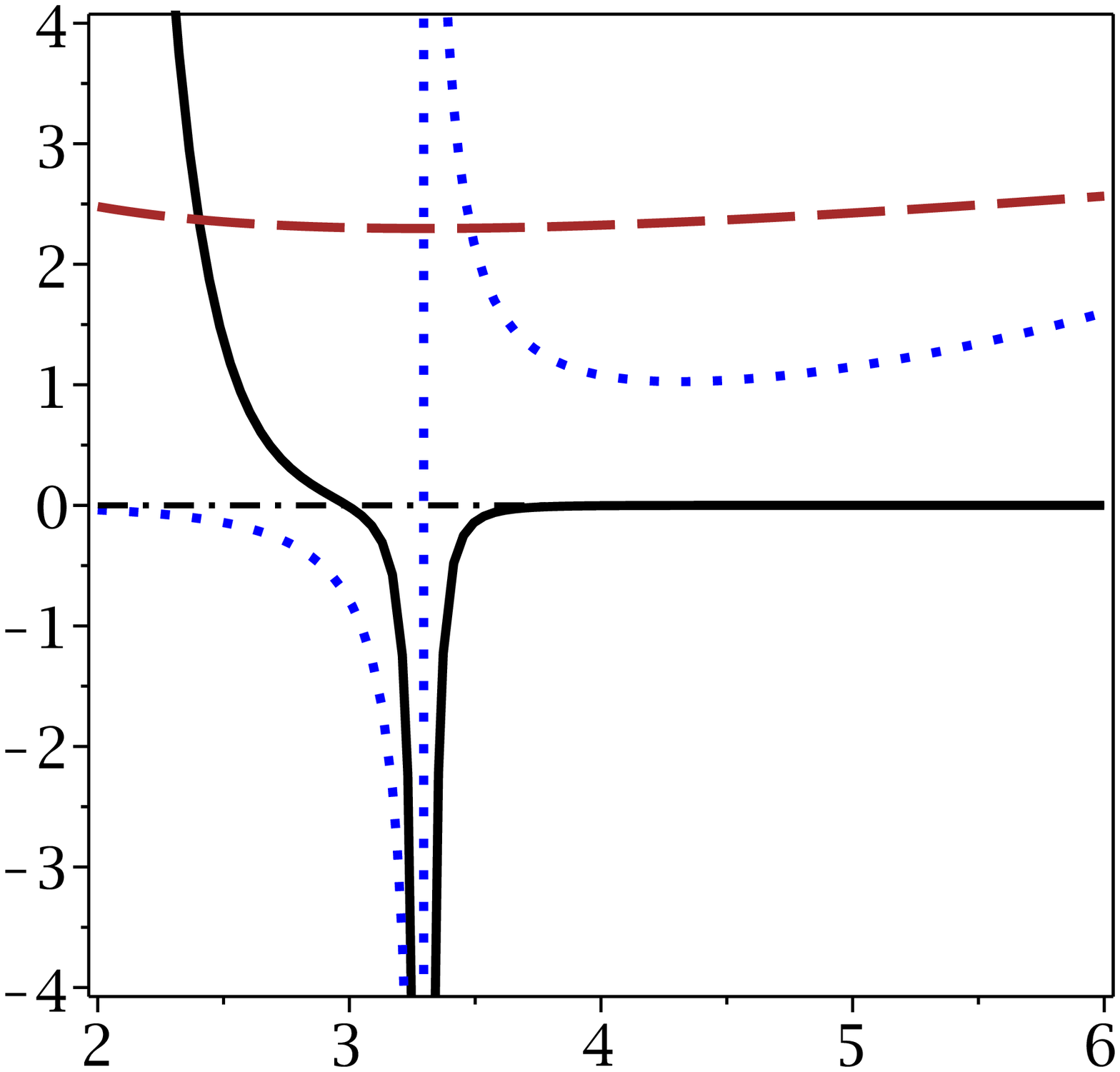}%
\end{array}
$%
\caption{For different scales: $\mathcal{R}$ (continuous line), $C_{Q}$
(dotted line) and $T$ (dashed line) versus $r_{+}$ for $\Lambda =-1$, $%
\protect\omega=1$, $c=1$ and $q=1$. (left diagram: $d=5$ \& middle
and right diagrams: $d=6$) } \label{Fig2}
\end{figure}

\begin{figure}[tbp]
$%
\begin{array}{ccc}
\epsfxsize=5cm \epsffile{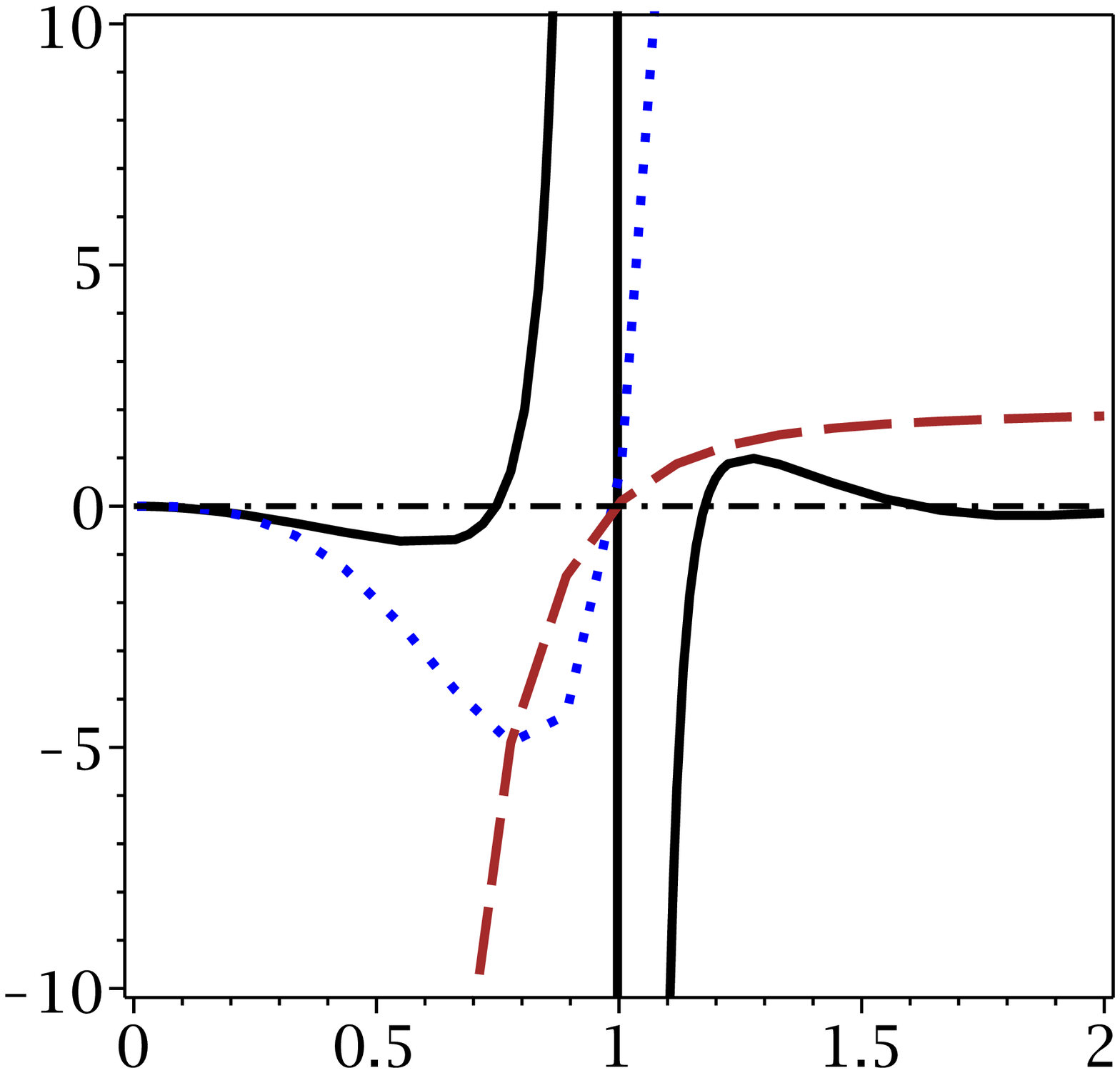} & \epsfxsize=5cm \epsffile{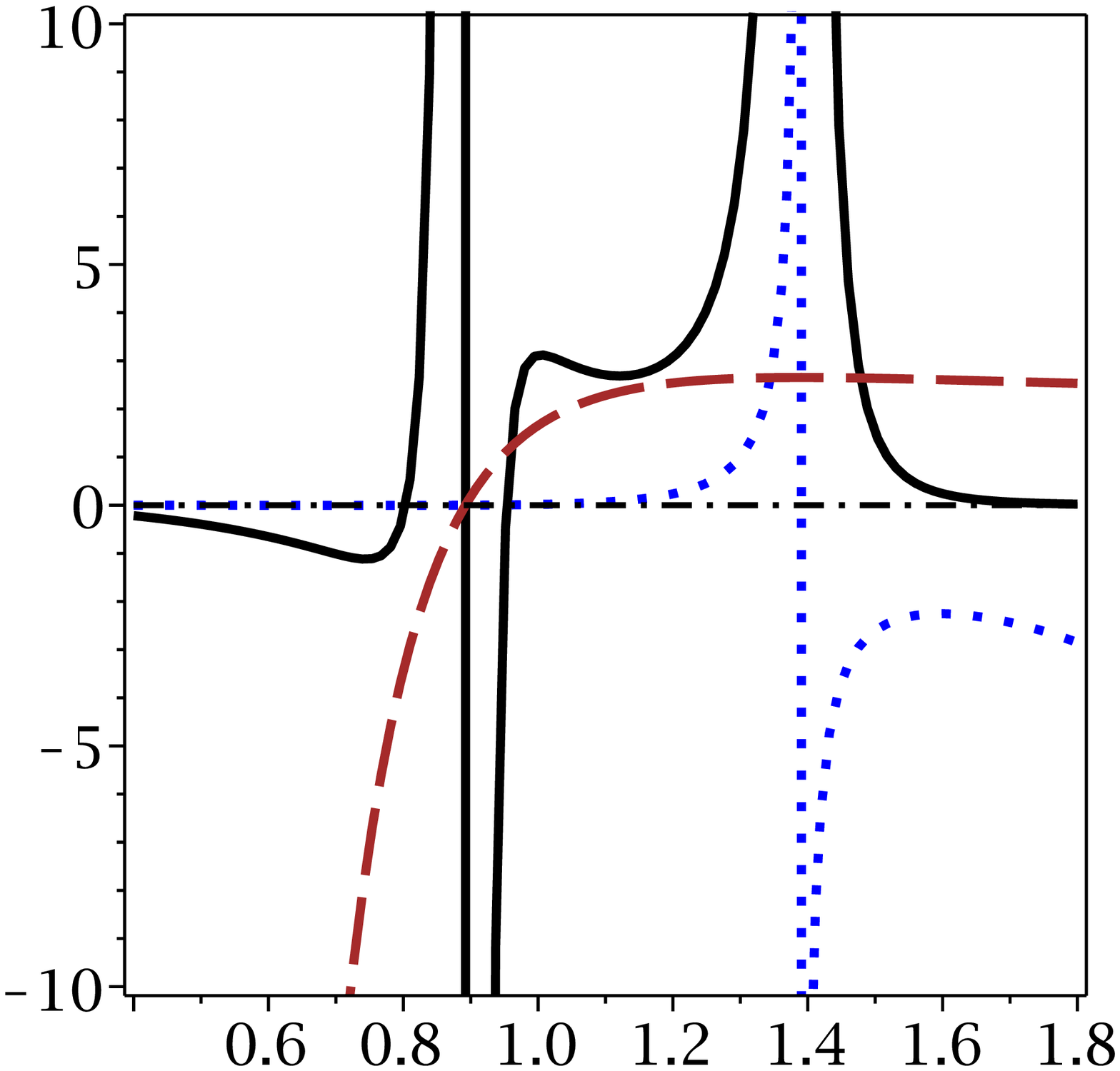} & %
\epsfxsize=5cm \epsffile{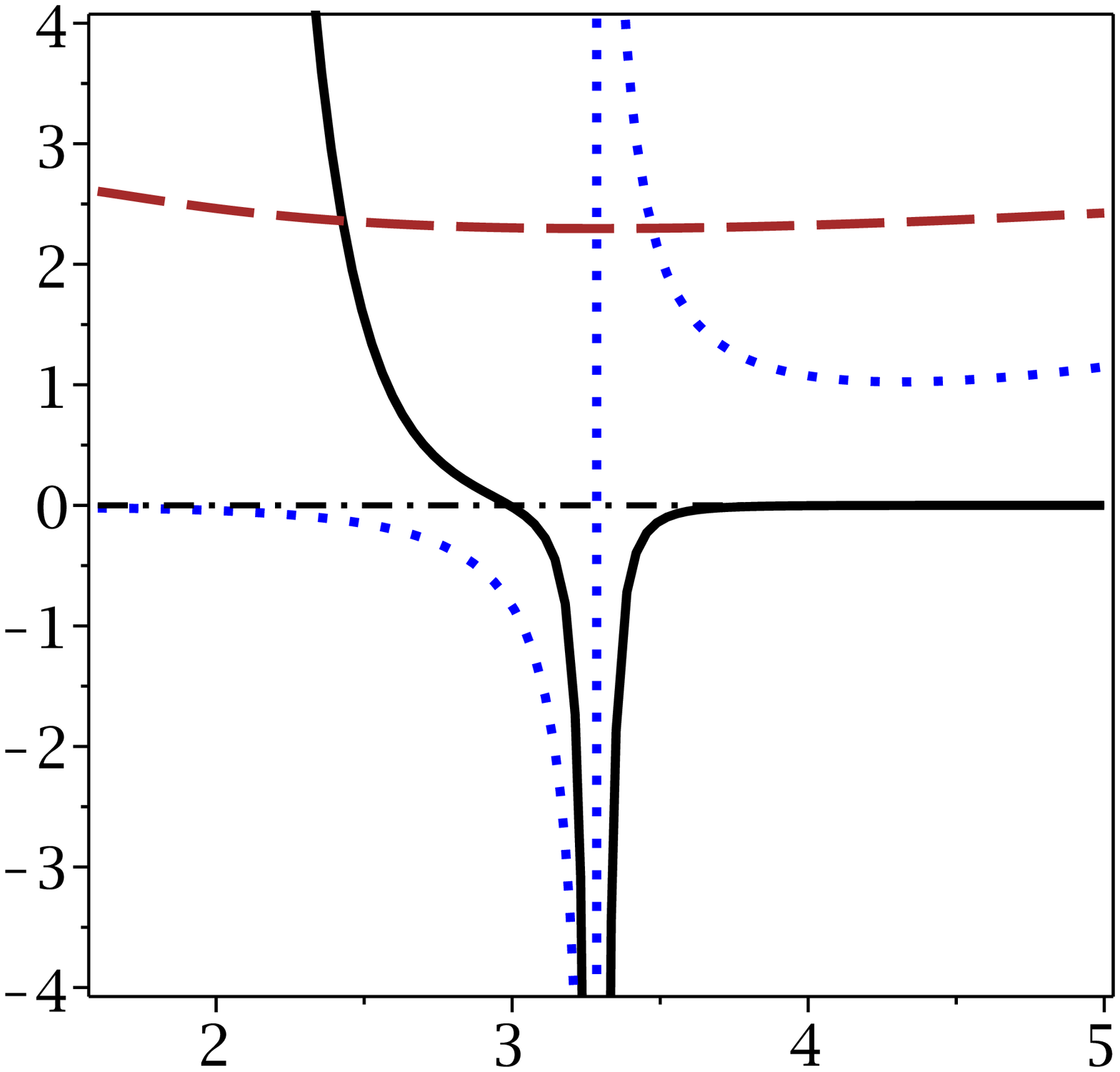}%
\end{array}
$%
\caption{For different scales: $\mathcal{R}$ (continuous line), $C_{Q}$
(dotted line) and $T$ (dashed line) versus $r_{+}$ for $\Lambda =-1$, $%
\protect\omega=1$, $c=1$ and $q=2$. (left diagram: $d=5$ \&
 middle and right diagrams: $d=6$) } \label{Fig3}
\end{figure}

\begin{figure}[tbp]
$%
\begin{array}{cc}
\epsfxsize=6cm \epsffile{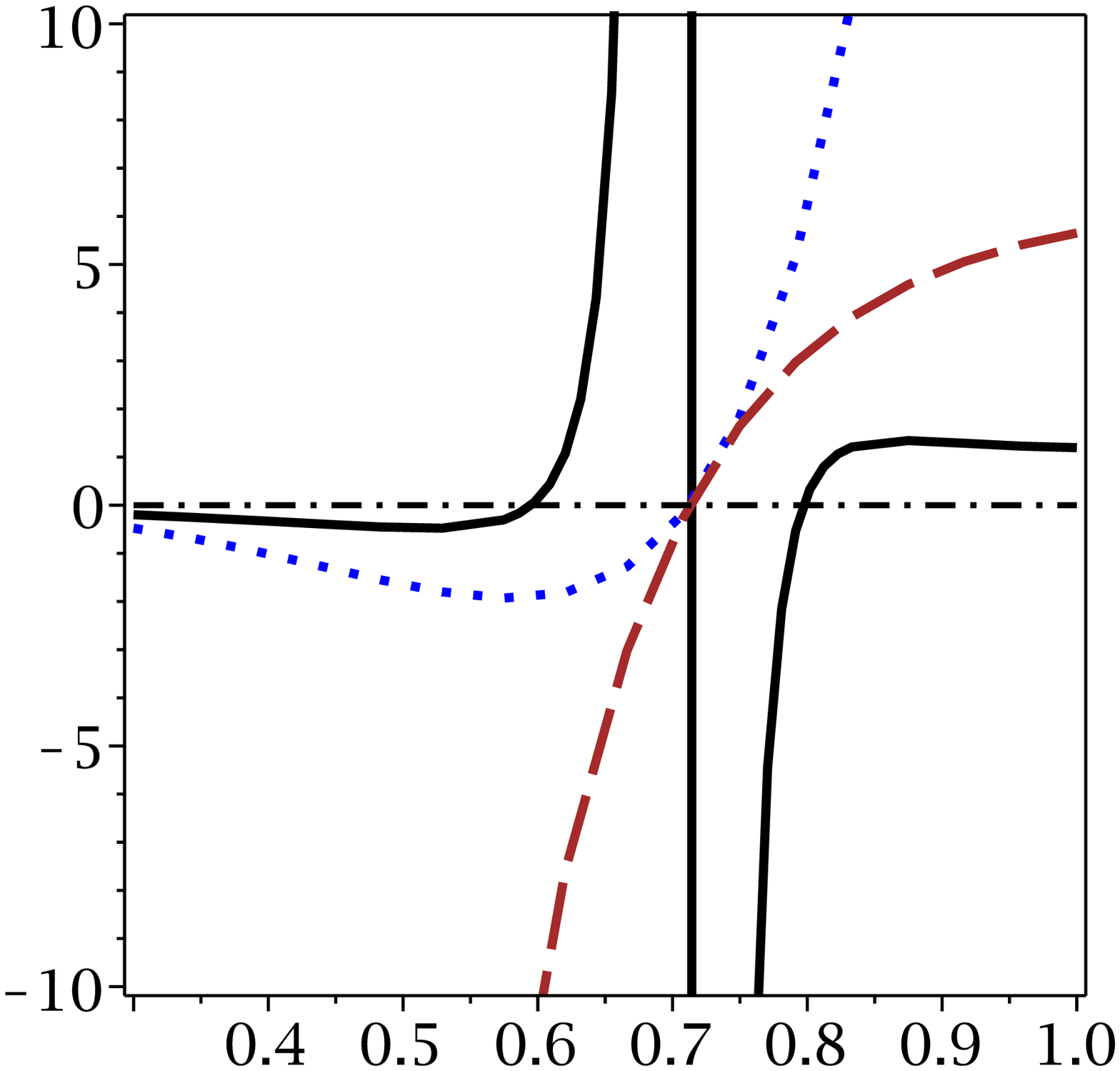} & \epsfxsize=6cm %
\epsffile{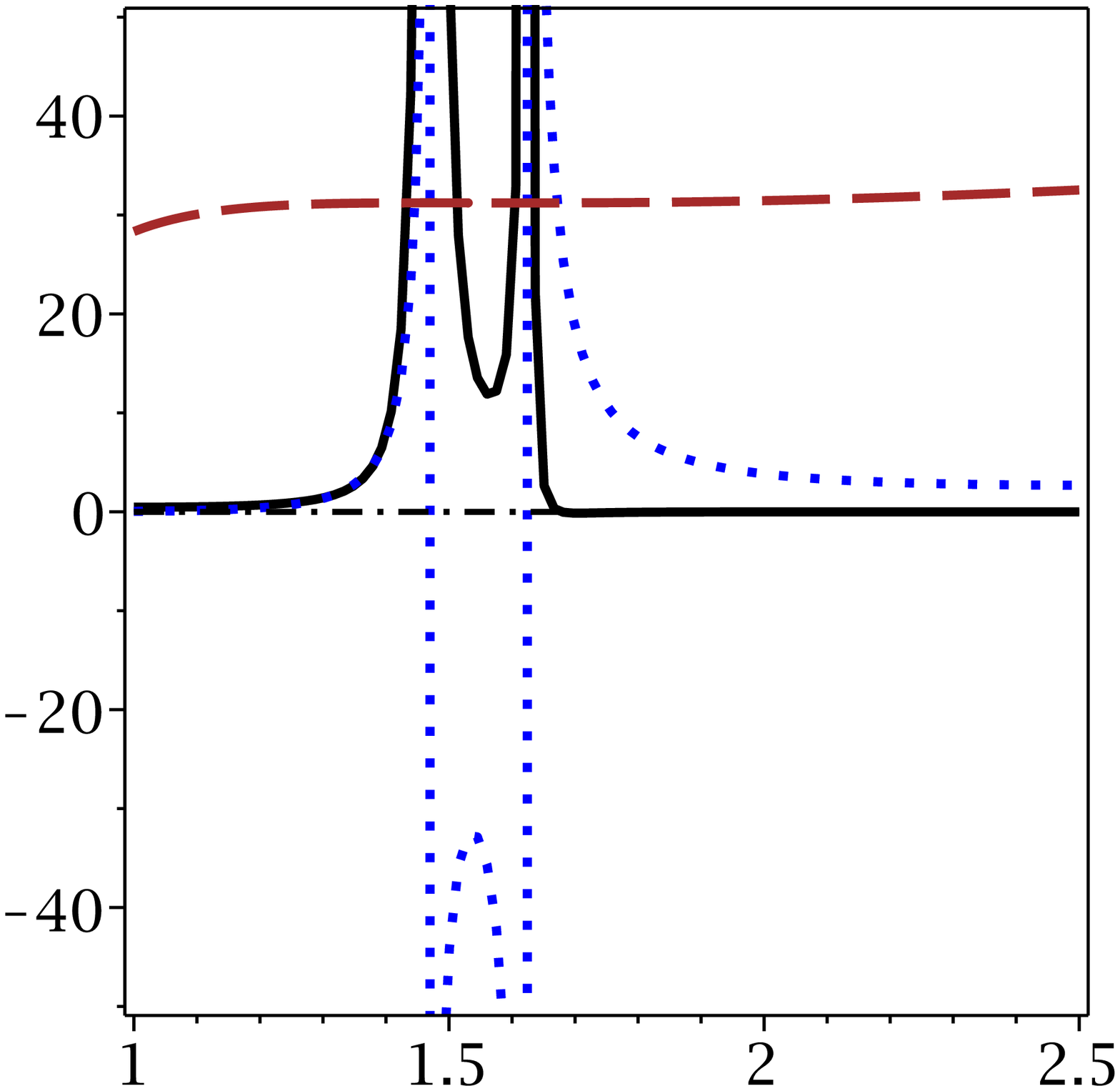} \\
\epsfxsize=6cm \epsffile{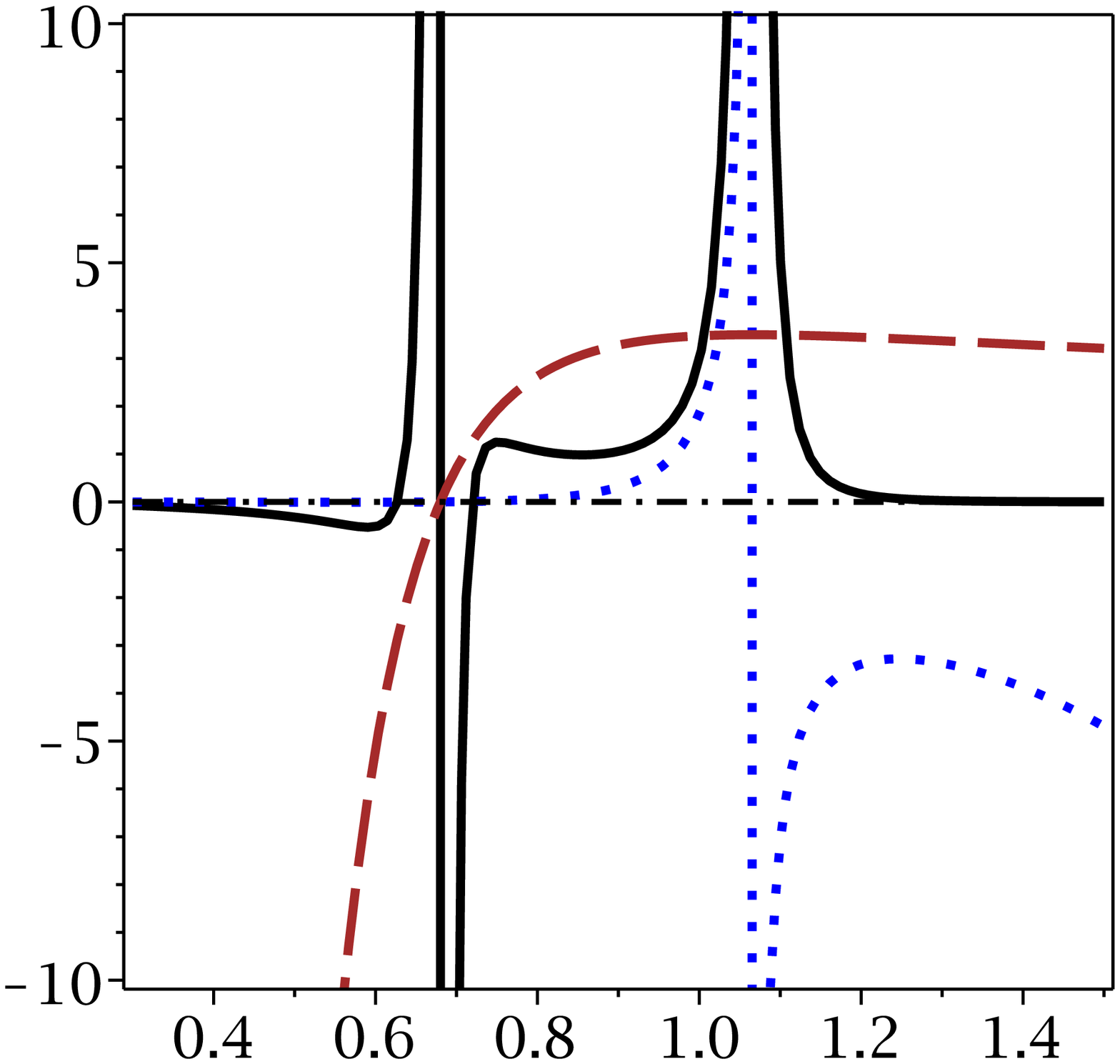} & \epsfxsize=6cm %
\epsffile{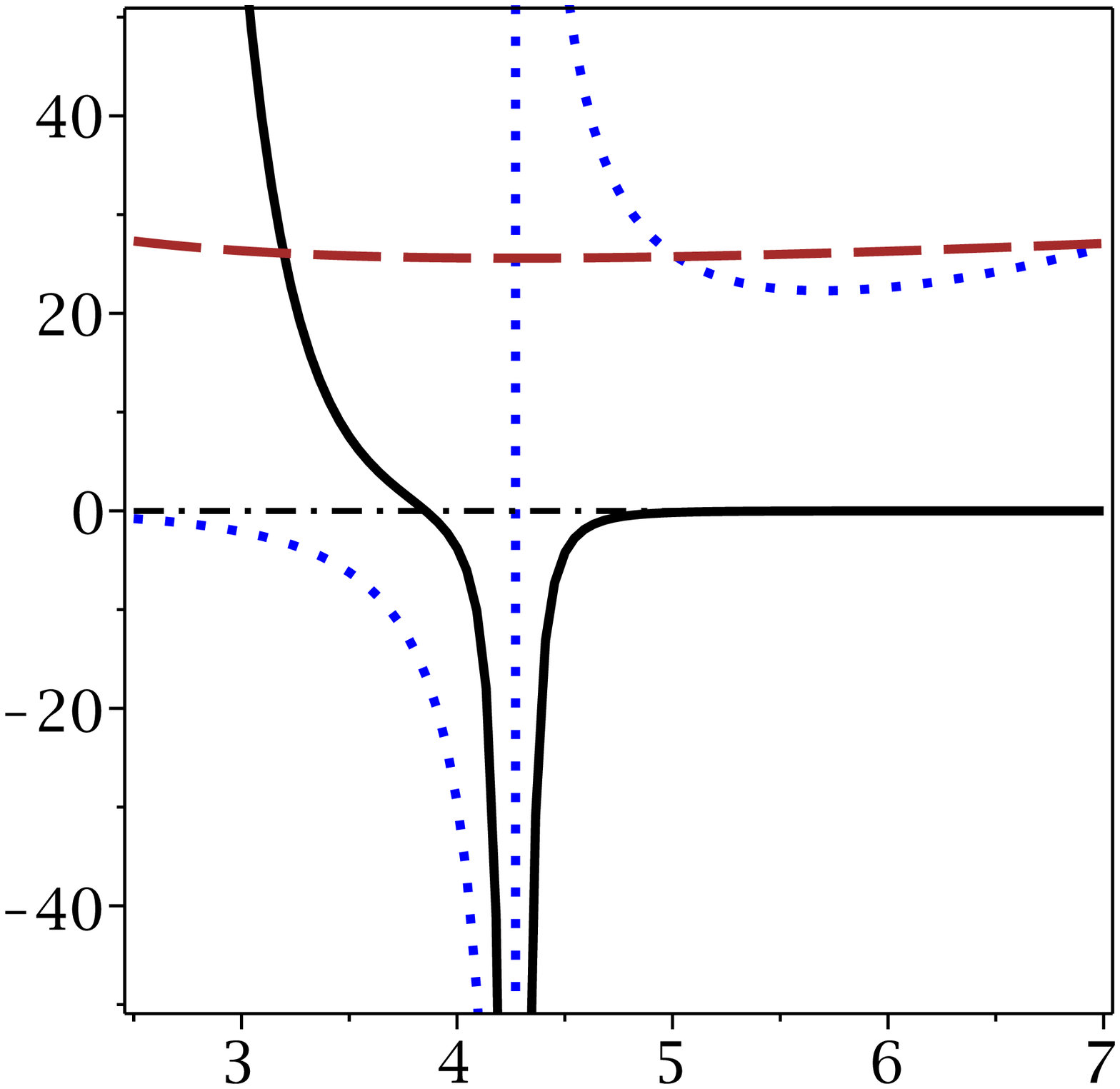}%
\end{array}
$%
\caption{For different scales: $\mathcal{R}$ (continuous line),
$C_{Q}$ (dotted line) and $T$ (dashed line) versus $r_{+}$ for
$\Lambda =-1$, $q=1$, $c=1$ and $\protect\omega=0.1$. (up
diagrams: $d=5$ \& down diagrams: $d=6$) } \label{Fig4}
\end{figure}

\begin{figure}[tbp]
$%
\begin{array}{ccc}
\epsfxsize=5cm \epsffile{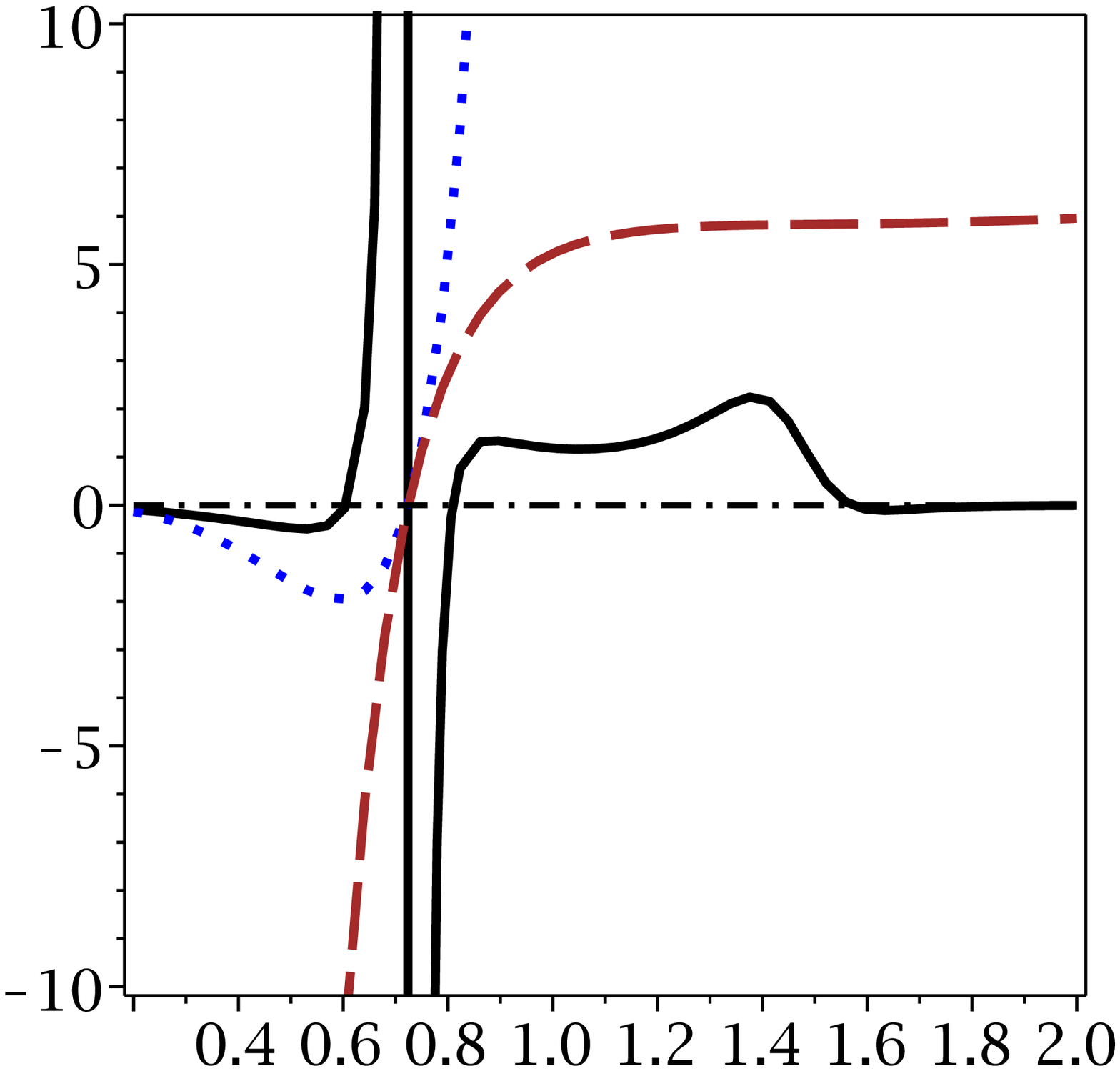} & \epsfxsize=5cm %
\epsffile{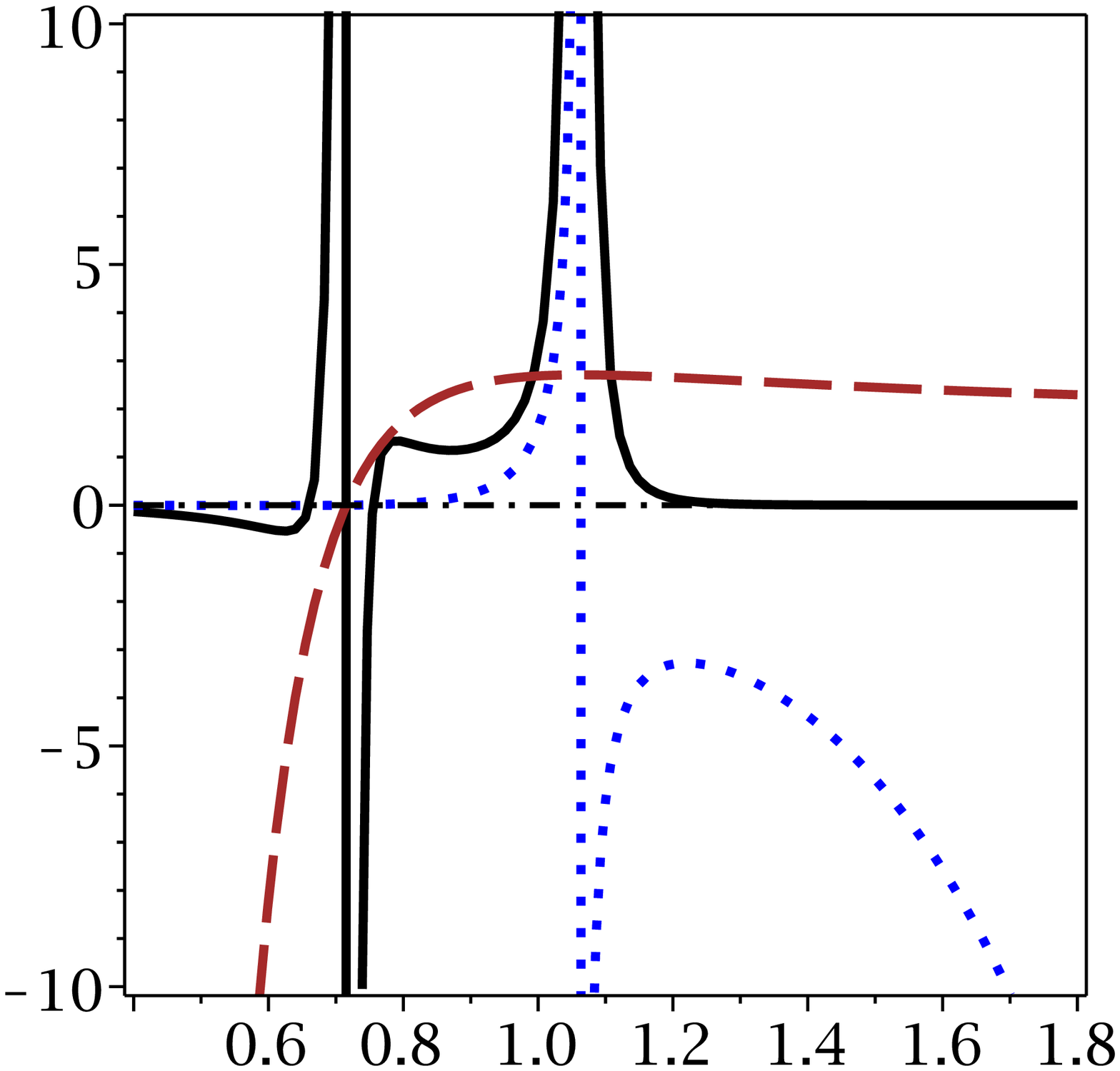} & \epsfxsize=5cm \epsffile{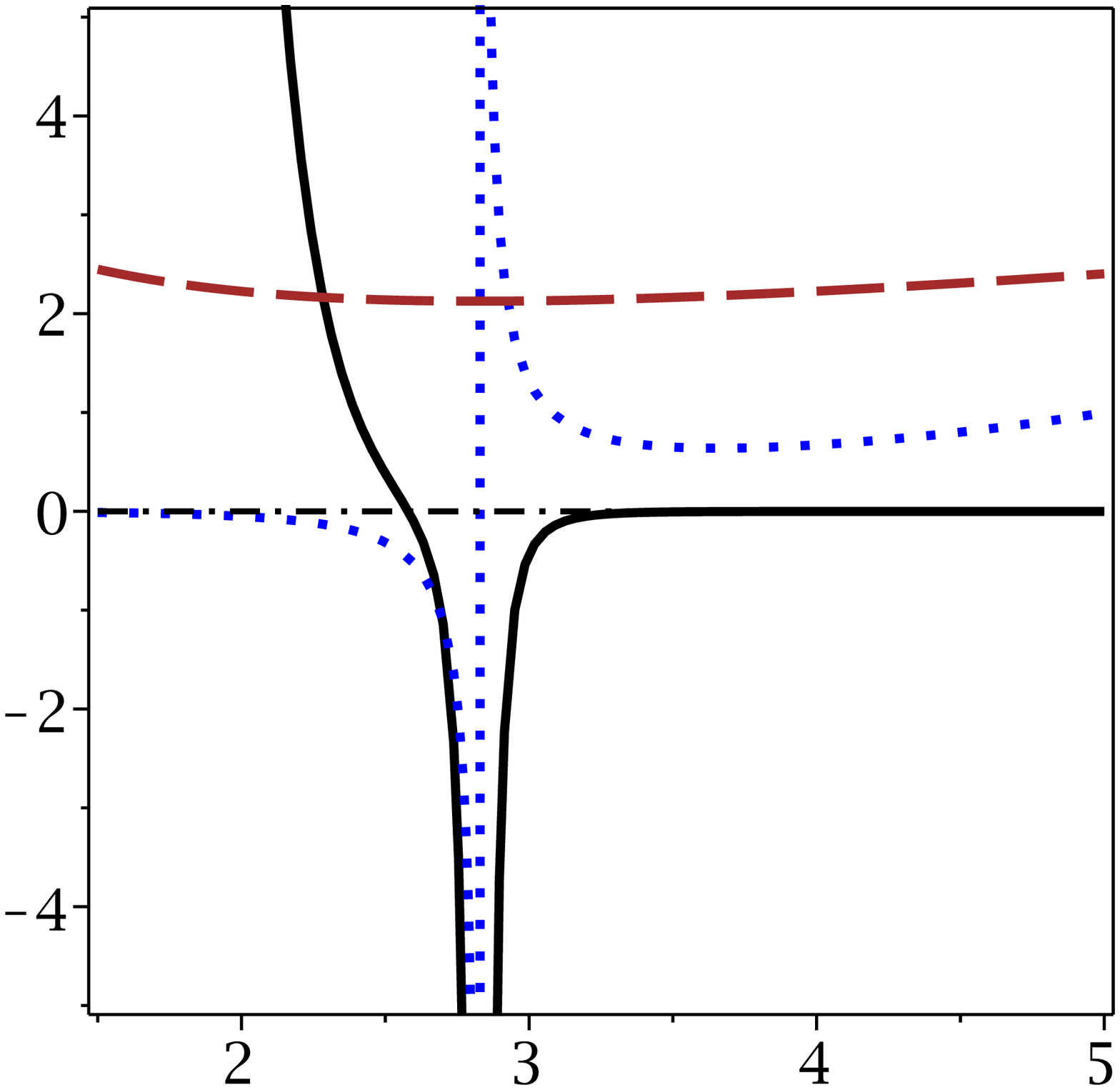}%
\end{array}
$%
\caption{For different scales: $\mathcal{R}$ (continuous line),
$C_{Q}$ (dotted line) and $T$ (dashed line) versus $r_{+}$ for
$\Lambda =-1$, $q=1$, $c=1$ and $\protect\omega=3$. (left diagram:
$d=5$ \& middle and right diagrams: $d=6$) } \label{Fig5}
\end{figure}

It is worthwhile to mention that for the case of physical limitation point,
numerical evaluation shows that there is only one root for this case which
will be seen by plotted graphs for the heat capacity. In the case of phase
transition point, interestingly, numerical evaluation shows that two cases
might happen: in one of these cases, there is no real root for Eq. (\ref%
{dive}), hence there is no phase transition for the black holes. Whereas in
the other case, due to the structure of Eq. (\ref{dive}), there will be two
roots for this equation which indicate the existence of two divergence
points, and therefore, could be interpreted as phase transition for these
black holes. This will be seen in plotted graphs (Figs. \ref{Fig1}-\ref{Fig5}%
) in more details.

Another approach for studying the phase transition of black holes
is through geometrical thermodynamics. There are several metrics
that one can employ in order to build geometrical phase space by
thermodynamical quantities. The well known ones are Ruppeiner,
Weinhold and Quevedo. It was previously argued that these metrics
may not provide us a completely flawless mechanism for studying
geometrical thermodynamics of specific types of black holes
\cite{NewMetric}. In this paper, we will show that the method of
geometrical thermodynamics reported in \cite{Niu} is not suitable
in the presence of scalar field. Recently, a new metric (HPEM
metric) was proposed in order to solve the problems that other
metrics may confront with \cite{NewMetric}.

According to Ref. \cite{Q1}, it is possible to derive, in
principle, an infinite number of Legendre invariant metrics. In
addition, it was shown that one of the simplest ways for obtaining
the Legendre invariant metrics is to apply a conformal
transformation. Comparing HPEM metric with Quevedo's one, we find
that they are the same up to a conformal factor. In addition,
despite Weinhold and Ruppeiner, both HPEM and Quevedo metrics
enjoy the same ($- + + + ...$) signature. Regarding the same
signature with difference in conformal factor, it is expected that
HPEM enjoys Legendre invariancy with different Legendre
multiplier.

On the other hand, we should mention an unavoidable comment
related to Legendre invariancy. Although it was thought that the
Legendre invariancy guarantees unique description of
thermodynamical metrics, it was shown that such invariancy alone
is not sufficient for the mentioned guarantee in terms of
thermodynamical curvatures \cite{Q2}. However, it was proven that
in addition to Legendre invariancy, we need to demand curvature
invariancy in various representations. Therefore, both Legendre
and curvature invariancies should be checked. In addition, there
are two issues with a fundamental relation between them; (I)
agreement of thermodynamical curvature results with usual
thermodynamical approaches (such as the heat capacity); (II)
curvature invariancy in addition to the Legendre invariancy. It is
important to probe the fundamentality of cases (I) and (II) for
discovering that regarding which case may lead to satisfy another
one. Although first case has been investigated for special cases
\cite{Q2}, the second one has been remained unanalyzed yet. One
may address them in an independent work in the future.

The Weinhold, Ruppeiner and Quevedo thermodynamical metrics are
given by
\begin{equation}
ds_{W}^{2}=Mg_{ab}^{W}dX^{a}dX^{b},  \label{Wein}
\end{equation}
\begin{equation}
ds_{R}^{2}=-MTg_{ab}^{W}dX^{a}dX^{b},  \label{Rupp2}
\end{equation}
\begin{equation}
ds_{Q}^{2}=\left( SM_{S}+QM_{Q}\right) \left(
-M_{SS}dS^{2}+M_{QQ}dQ^{2}\right) ,  \label{Quevedo}
\end{equation}
where $g_{ab}^{W}=\partial ^{2}M\left( X^{c}\right) /\partial
X^{a}\partial X^{b}$ and also $X^{a}\equiv X^{a}\left(
S,N^{i}\right)$, in which $N^{i}$ denotes other extensive
variables of the system. The HPEM metric, with $2$ extensive
parameters (entropy and electric charge), is
\begin{equation}
ds_{HPEM}^{2}=\frac{SM_{S}}{M_{QQ}^{3}}\left(
-M_{SS}dS^{2}+M_{QQ}dQ^{2}\right) , \label{NewMetric}
\end{equation}%
where $M_{X}=\left( \frac{\partial M}{\partial X}\right) $ and
$M_{XX}=\left( \frac{\partial ^{2}M}{\partial X^{2}}\right) $.
Calculations show that the numerator and denominator of HPEM
method for this type of thermodynamical system are given as
\cite{NewMetric}
\begin{eqnarray}
num(\mathcal{R}_{HPEM})
&=&6S^{2}M_{S}^{2}M_{QQ}M_{SS}^{2}M_{QQQQ}-6SM_{S}^{2}M_{QQ}^{2}M_{SS}M_{SSQQ}+2SM_{SQQ}^{2}M_{S}^{2}M_{QQ}M_{SS}
\nonumber \\
&&+2\left[ SM_{S}M_{SSS}-\frac{1}{2}M_{SS}\left( SM_{SS}-M_{S}\right) \right]
SM_{QQ}^{2}M_{S}M_{SQQ}-9S^{2}M_{QQQ}^{2}M_{S}^{2}M_{SS}^{2}  \nonumber \\
&&+4\left[ \frac{1}{4}M_{SQ}M_{SS}+M_{S}M_{SSQ}\right]
S^{2}M_{QQ}M_{S}M_{QQQ}+\left[
S^{2}M_{S}^{2}M_{SSQ}-S^{2}M_{SQ}M_{SS}M_{S}M_{SSQ}\right.  \nonumber \\
&&\left. SM_{QQ}M_{S}\left( SM_{SS}-M_{S}\right) M_{SS}-2\left(
S^{2}M_{SS}^{3}+M_{S}^{2}M_{SS}\right) M_{QQ}+2S^{2}M_{SQ}^{2}M_{SS}^{2}
\right] M_{QQ}^{2},  \label{NumNew}
\end{eqnarray}%
and
\begin{equation}
denom(\mathcal{R}_{HPEM})=S^{3}M_{S}^{3}M_{SS}^{2}.  \label{denNew}
\end{equation}

It is obvious that in general, the roots of the numerator and
denominator of this Ricci scalar are not coinciding. Meaning that
the case, where the divergency of Ricci scalar could be cancelled
by its root, does not take place. The denominator of Ricci scalar
of HPEM metric contains numerator and denominator of the heat
capacity. In other words, divergence points of the Ricci scalar of
HPEM metric coincide with both roots and phase transition points
of the heat capacity. Therefore, all the phase transition and
limitation points are included in the divergencies of Ricci scalar
of this metric.

In what follows, we study the stability and phase transition of the charged
BD black holes in the context of canonical ensemble by calculating heat
capacity. Then, we study geometrical thermodynamics of these black holes by
using the HPEM thermodynamical metric. Before conducting the research in the
context of HPEM method, we will show that Weinhold, Ruppeiner and Quevedo
metrics fail to provide effective results.

First of all, one should take this fact into consideration that the sign of
temperature is putting a restriction of system being physical or
non-physical. In other words, the negativity of temperature denotes a
non-physical system which in our case is black hole. It is evident that,
there is a critical horizon radius, say $r_{c}$, in which for $r_{+}<r_{c}$
the temperature of the system is negative. Therefore, in this region
solutions are non-physical. Since for $r_{+}>r_{c}$ system has positive
temperature, the horizon radius of physical black holes are placed in this
region.

In case of $5$--dimensional black holes, for sufficiently small values of
electric charge (Fig. \ref{Fig1} top), BD coupling constant (Fig. \ref{Fig4}
top), these black holes have three characteristic points. One is related to
the root of heat capacity and the others are related to its divergencies ($%
r_{Div1}$ and $r_{Div2}$ in which $r_{Div1}<r_{Div2}$). It is clear that the
root of heat capacity and $r_{c}$ are the same and for the case $r_{+}<r_{c}$
the heat capacity is negative but due to negativity of temperature, the
system is not physical. On the other hand, for the region $%
r_{c}<r_{+}<r_{Div1}$, heat capacity is positive definite. Therefore, in
this region, the system is in a stable state. As for the $%
r_{Div1}<r_{+}<r_{Div2}$ the heat capacity has a negative value which is
denoted as instability of the black holes. In other words, in case of $%
r_{+}=r_{Div1}$ system may go under phase transition from large and unstable
black hole to a smaller and stable one. In case of $r_{+}=r_{Div2}$ system
will have another phase transition and its stability will change from
unstable to stable one. In other words, for $r_{+}>r_{Div2}$ the heat
capacity has a positive value and the black hole is stable. It is notable
that for sufficiently large values of $q$ (Figs. \ref{Fig2} and \ref{Fig3}
left), $\omega $ (Fig. \ref{Fig5} left), divergencies of the heat capacity
vanish and the black hole is thermally stable for $r_{+}>r_{c}$, without any
phase transition.

In case of $6$--dimensional charged BD black holes, we find the following
results. It is clear that in case of $6$--dimensional, using considered
values in case of $5$--dimensional solutions, in all cases one root and two
divergence points for heat capacity are observed. In these cases, heat
capacity and phase transitions have similar behavior as what was observed in
$5$--dimensions charged BD black holes. $r_{Div1}$ is an increasing function
of electric charge (Figs. \ref{Fig2} and \ref{Fig3} middle) while $r_{Div2}$
is a decreasing function of BD coupling constant (Figs. \ref{Fig2} and \ref%
{Fig5} right). We should note that the effects of variation of $q$ is so
small on second divergence point (Figs. \ref{Fig2} and \ref{Fig3} right).

\begin{figure}[tbp]
$%
\begin{array}{cc}
\epsfxsize=6cm \epsffile{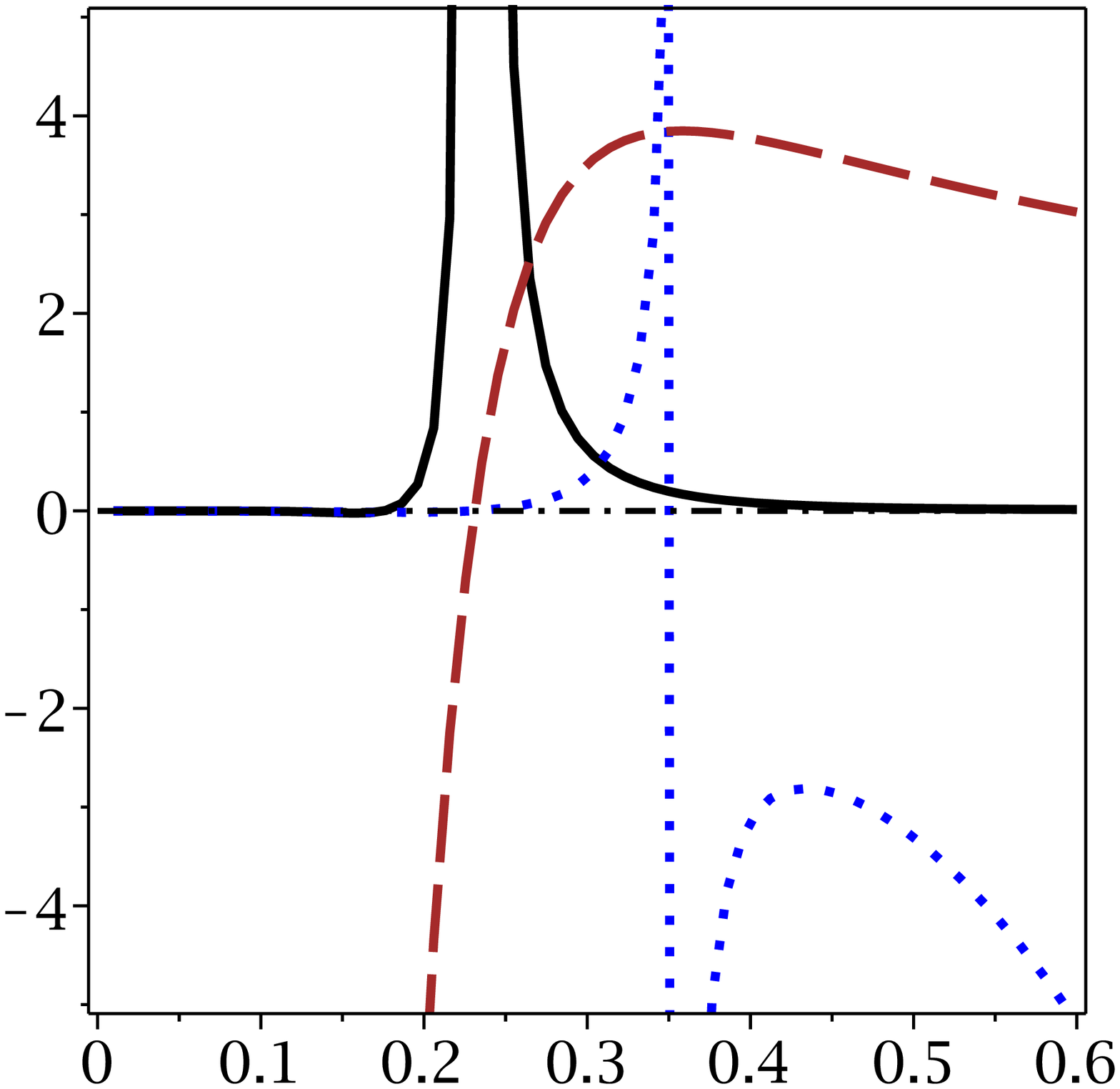} & \epsfxsize=6cm \epsffile{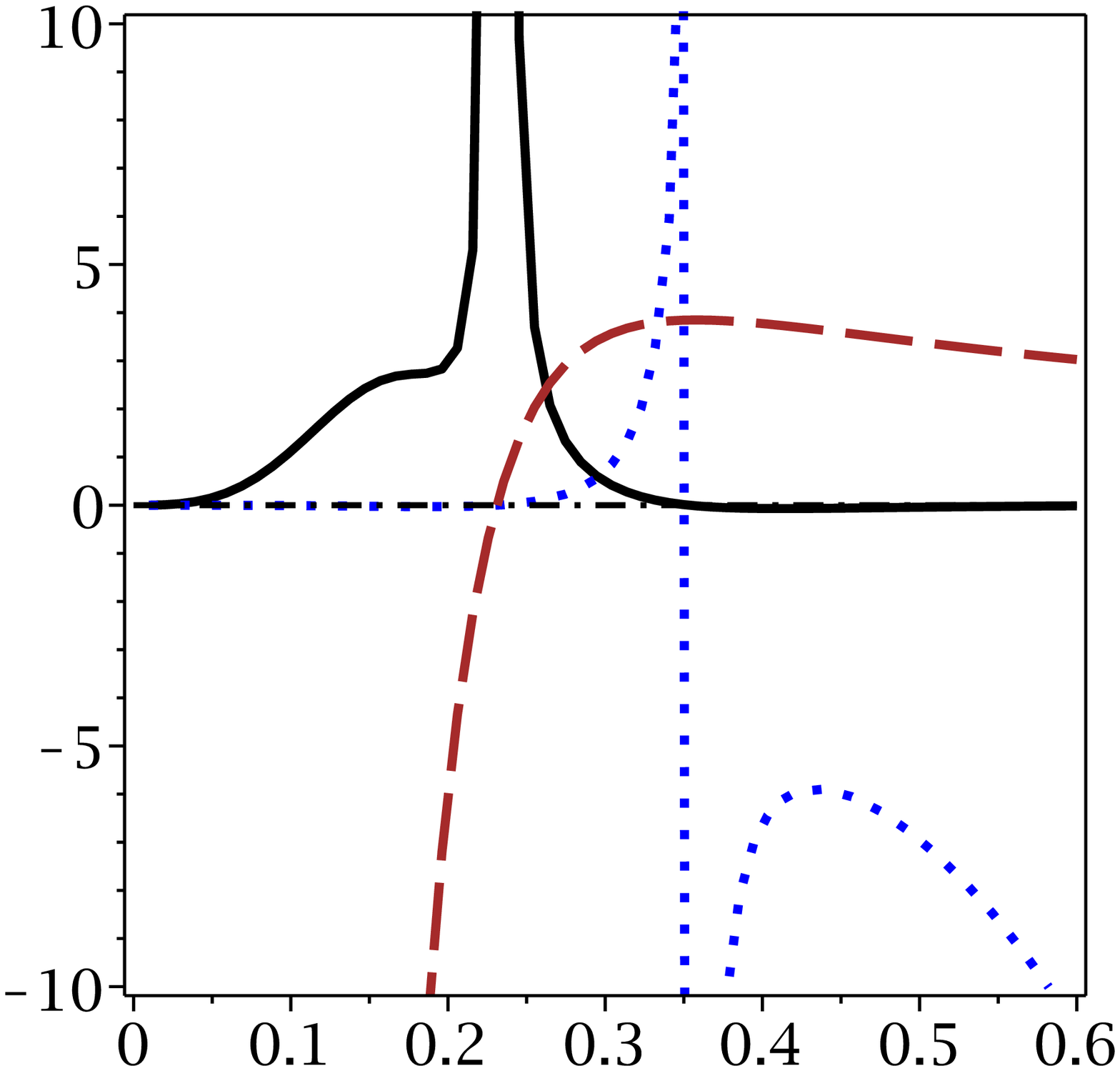}
\\
\epsfxsize=6cm \epsffile{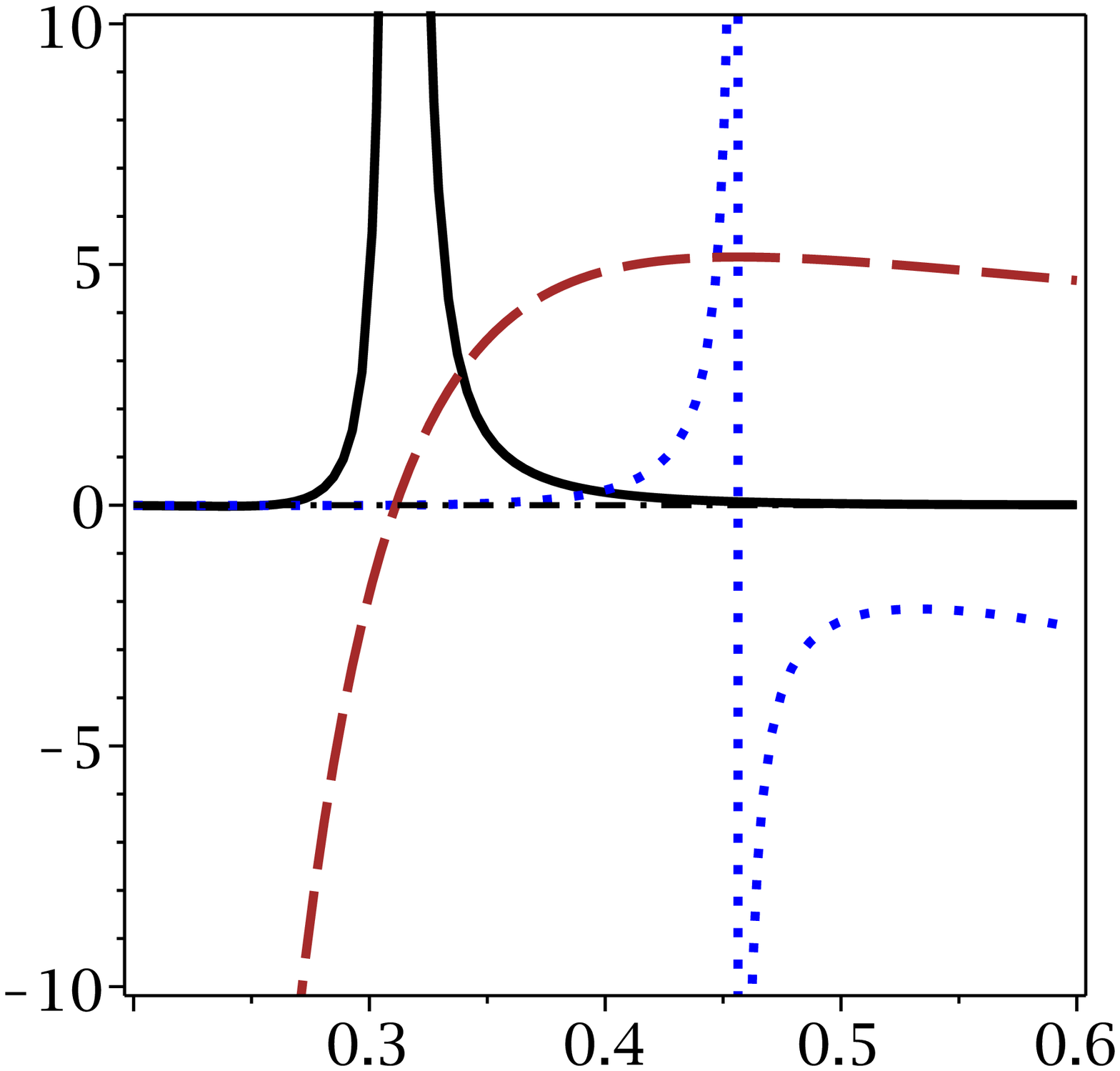} & \epsfxsize=6cm \epsffile{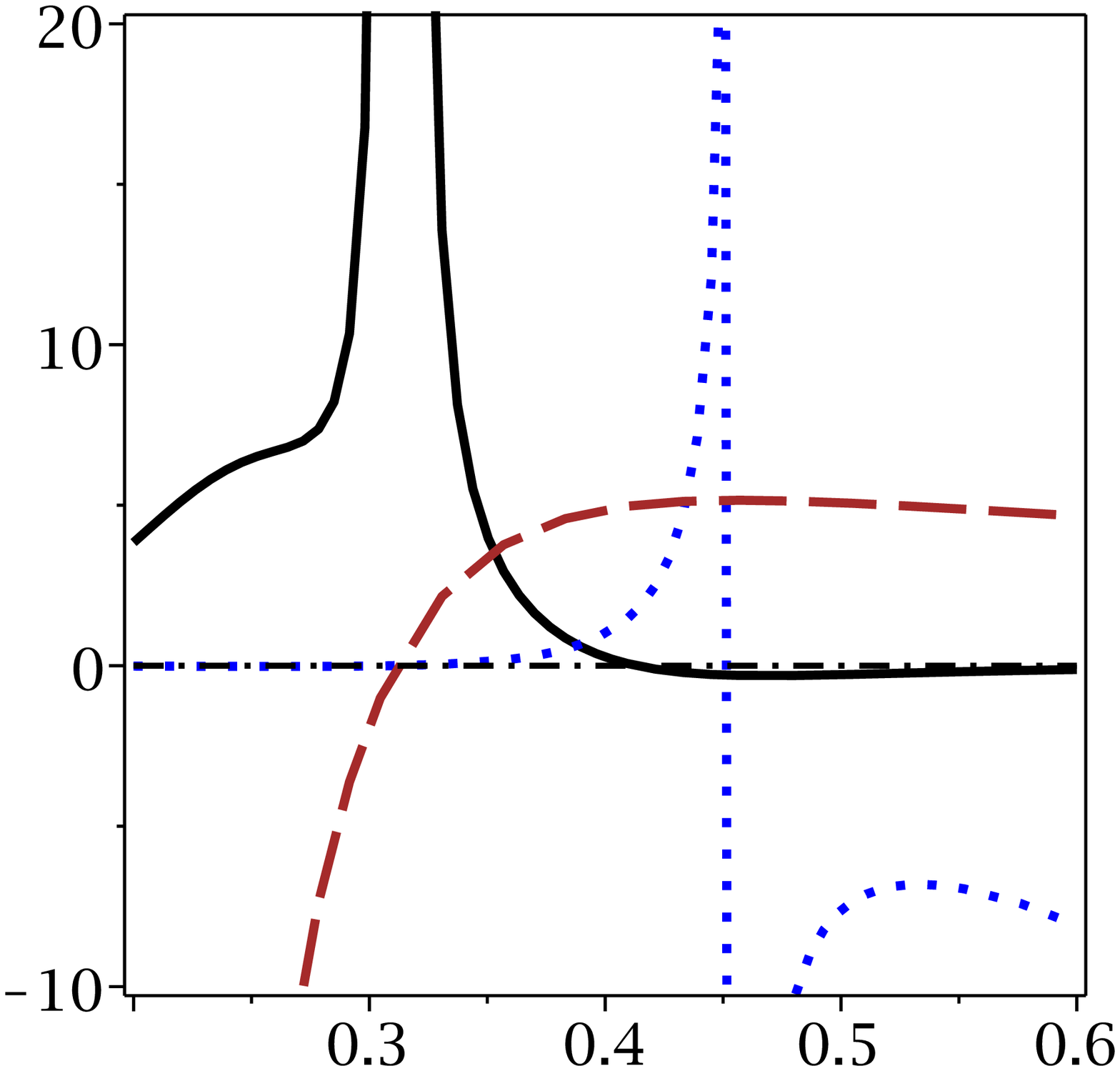}%
\end{array}
$%
\caption{Left panel: Weinhold, right panel: Ruppeiner;
$\mathcal{R}$ (continuous line), $C_{Q}$ (dotted line) and $T$
(dashed line) versus $r_{+}$ for $\Lambda =-1$,
$\protect\omega=1$, $c=1$ and $q=0.1$. (up diagrams: $d=5$ \& down
diagrams: $d=6$) } \label{Fig11}
\end{figure}


Next, by employing the metrics that were defined, Eqs. (\ref{Wein})-(\ref%
{NewMetric}), we construct a thermodynamical spacetime with mass as
thermodynamical potential. Using Eqs. (\ref{m})--(\ref{Q}) one can write
total mass of the black holes as a function of extensive parameters. By
employing Eqs. (\ref{Wein})-(\ref{NewMetric}) the spacetime will be
constructed. Next, we calculate the curvature scalar of the mentioned
thermodynamical metrics. Analytical calculations of the curvature scalar is
too large and, therefore, we leave out the analytical result for reasons of
economy.

Studying Fig. \ref{Fig11} shows that using Weinhold and Ruppeiner leads to a
mismatch between divergency of the Ricci scalar of these two metrics and
divergency of the heat capacity. In other words, the phase transitions of
the Ruppeiner and Weinhold metrics do not coincide with phase transitions of
the canonical ensemble, hence, heat capacity. Next, we present the behavior
of obtained results for HPEM method in plotted graphs (Figs. \ref{Fig1}-\ref%
{Fig5}). These figures show that divergencies of thermodynamical
Ricci scalar coincide with root and divergence points of the heat
capacity. As for Quevedo metric, we will follow another approach
to show that it fails to produce suitable results. It is a matter
of calculation to show that using Quevedo metric (\ref{Quevedo}),
one can find following denominator for its Ricci scalar
\begin{equation}
denom(\mathcal{R}_{Q})=\left( SM_{S}+QM_{Q}\right)
^{3}M_{SS}^{2}M_{QQ}^{2}. \label{RQ1}
\end{equation}

Here, three terms contribute to divergencies of the Ricci scalar
of Quevedo metric. $M_{SS}$ ensures that phase transitions of the
heat capacity and some of the divergencies of the Ricci scalar of
this metric coincide, where the other two terms will result into
extra divergencies that are not coincidence with any phase
transition. In other words, the results of using this metric are
not consistent with the results of heat capacity. Therefore, if
one uses this metric independent of heat capacity, due to being
plugged with extra divergencies, it is not possible to make
acceptable statements regarding the physical properties of the
system. For further clarification, we give an example. For $d=5$
and $\Lambda=\omega=c=1$, one can find the following relation for
the Ricci scalar of Quevedo metric
\begin{eqnarray}
R=&&\frac{-104799744 \pi^{2}\;r_{+}^{\frac{228}{29}}}{841\;\Gamma
_{1}^{2}\;\Gamma _{2}^{3}}\left\{ 11981115\;r_{+}^{\frac{390}{29}%
}-8575308\;r_{+}^{\frac{336}{29}}-83387808\;r_{+}^{\frac{282}{29}%
}+116784640\;r_{+}^{\frac{228}{29}}\right.  \nonumber \\
&&\left. -5880114\;r_{+}^{\frac{222}{29}}+33781860\;r_{+}^{\frac{168}{29}%
}+343570752\;r_{+}^{\frac{114}{29}}-123023853\;r_{+}^{\frac{54}{29}%
}-133929936\right\},
\end{eqnarray}
where $\Gamma_{1}$ term
\begin{equation}
\Gamma_{1}=9\;r_{+}^{\frac{168}{29}}+28\;r_{+}^{\frac{114}{29}}-47,
\end{equation}
in the denominator, ensures that all the divergencies of the heat
capacity are coincided with some of the divergencies of the Ricci
scalar, whereas, $\Gamma_{2}$ term
\begin{equation}
\Gamma _{2}=171\;r_{+}^{\frac{168}{29}}-532\;r_{+}^{\frac{114}{29}}-333,
\end{equation}%
provides extra divergencies, which are not matched with any phase
transition point. It is worthwhile to mention that the root of
numerator does not cancel extra divergencies of the Ricci scalar.
Therefore, this simple example shows that Quevedo metric is
plugged with extra divergencies which are not consistent with
phase transition points.

Therefore, HPEM metric provides a successful mechanism for studying the
places of root and phase transition points of these black holes in context
of canonical ensemble. It is worthwhile to mention that the behavior of the
thermodynamical Ricci scalar is different around root and divergence points
of the heat capacity. In other words, in case of divergency of Ricci scalar
coinciding with the root of heat capacity, the behavior of the diagrams
differs from the case in which divergency of the Ricci scalar coincides with
divergencies of heat capacity. Therefore, considering HPEM method, we find
that the root and phase transitions are distinguishable from one another. In
order to recognize physical limitation point ($r_{c}$) from phase transition
points, one can extract the following information from the figures. In case
of root of heat capacity ($S=S_{0}$), hence, physical limitation point,
there is a change in sign of Ricci scalar from $+\infty $ to $-\infty $.
Whereas for the case of the divergence points of the heat capacity, hence,
second order phase transitions, the sign of the Ricci scalar stays fixed.
Therefore, this change/unchange in sign is a characteristic that enables one
to distinguish root of heat capacity from phase transition points. We
summarize the mentioned information in table I.

\begin{center}
\begin{tabular}{c}
\begin{tabular}{ccc}
\hline\hline
& $C_{Q}$ & $\mathcal{R}$ \\ \hline\hline
$S=S_{0}$ & ${\lim_{S\rightarrow S_{0}}C_{Q}}=0$ & $%
\begin{array}{c}
{\lim_{S\rightarrow S_{0}^{-}}\mathcal{R}}=+\infty \\
{\lim_{S\rightarrow S_{0}^{+}}\mathcal{R}}=-\infty%
\end{array}%
$ \\ \hline
$S=S_{1c}$ & $%
\begin{array}{c}
{\lim_{S\rightarrow S_{1c}^{-}}C_{Q}}=+\infty \\
{\lim_{S\rightarrow S_{1c}^{+}}C_{Q}}=-\infty%
\end{array}%
$ & $%
\begin{array}{c}
{\lim_{S\rightarrow S_{\infty }^{-}}\mathcal{R}}=+\infty \\
{\lim_{S\rightarrow S_{\infty }^{+}}\mathcal{R}}=+\infty%
\end{array}%
$ \\ \hline
$S=S_{2c}>S_{1c}$ & $%
\begin{array}{c}
{\lim_{S\rightarrow S_{2c}^{-}}C_{Q}}=-\infty \\
{\lim_{S\rightarrow S_{2c}^{+}}C_{Q}}=+\infty%
\end{array}%
$ & $%
\begin{array}{c}
{\lim_{S\rightarrow S_{\infty }^{-}}\mathcal{R}}=-\infty \\
{\lim_{S\rightarrow S_{\infty }^{+}}\mathcal{R}}=-\infty%
\end{array}%
$ \\ \hline
\end{tabular}
\\
Table $I$: comparison of root with phase transition points%
\end{tabular}
\end{center}

\begin{figure}[tbp]
$%
\begin{array}{cc}
\epsfxsize=6cm \epsffile{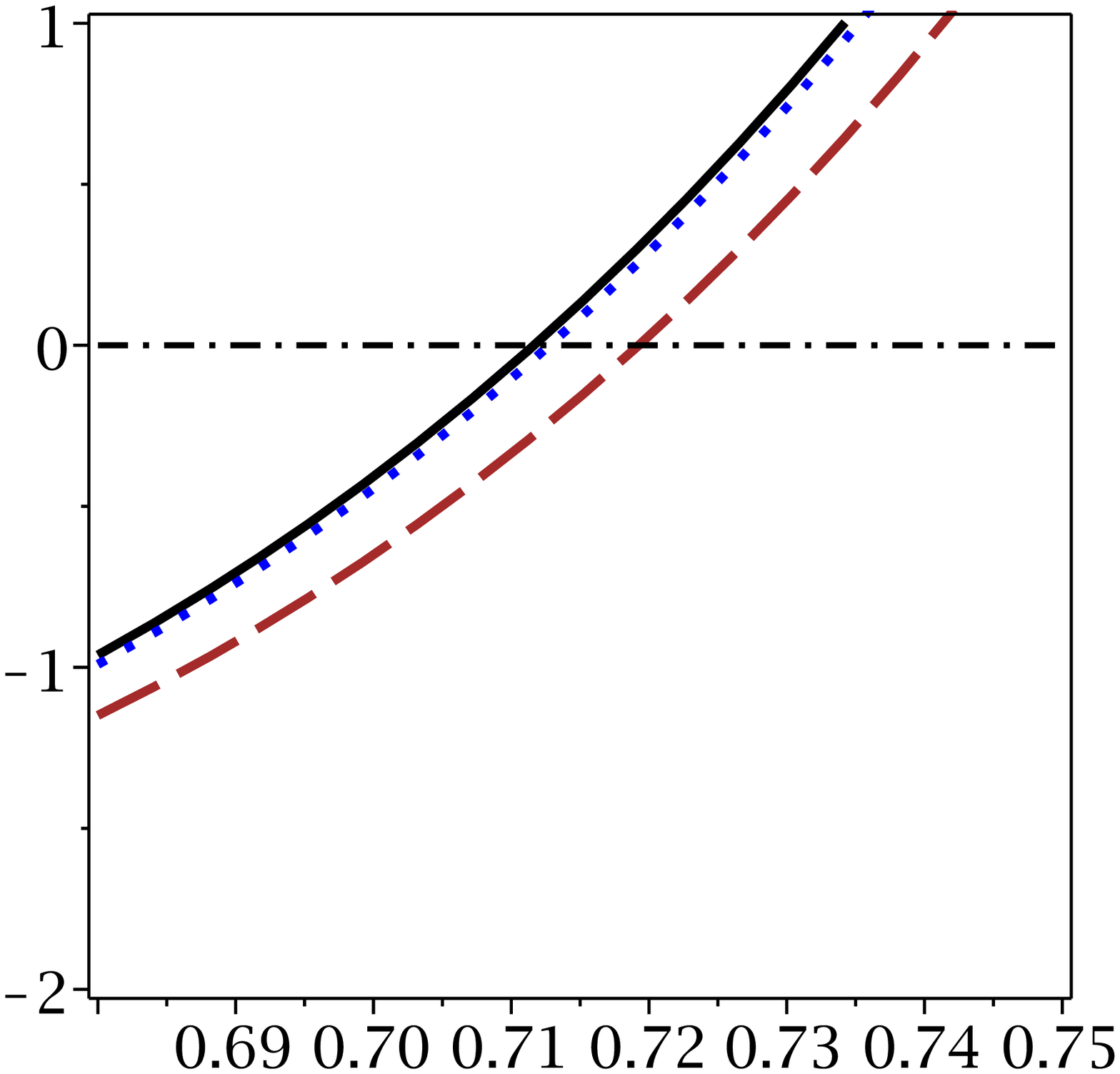} & \epsfxsize=6cm %
\epsffile{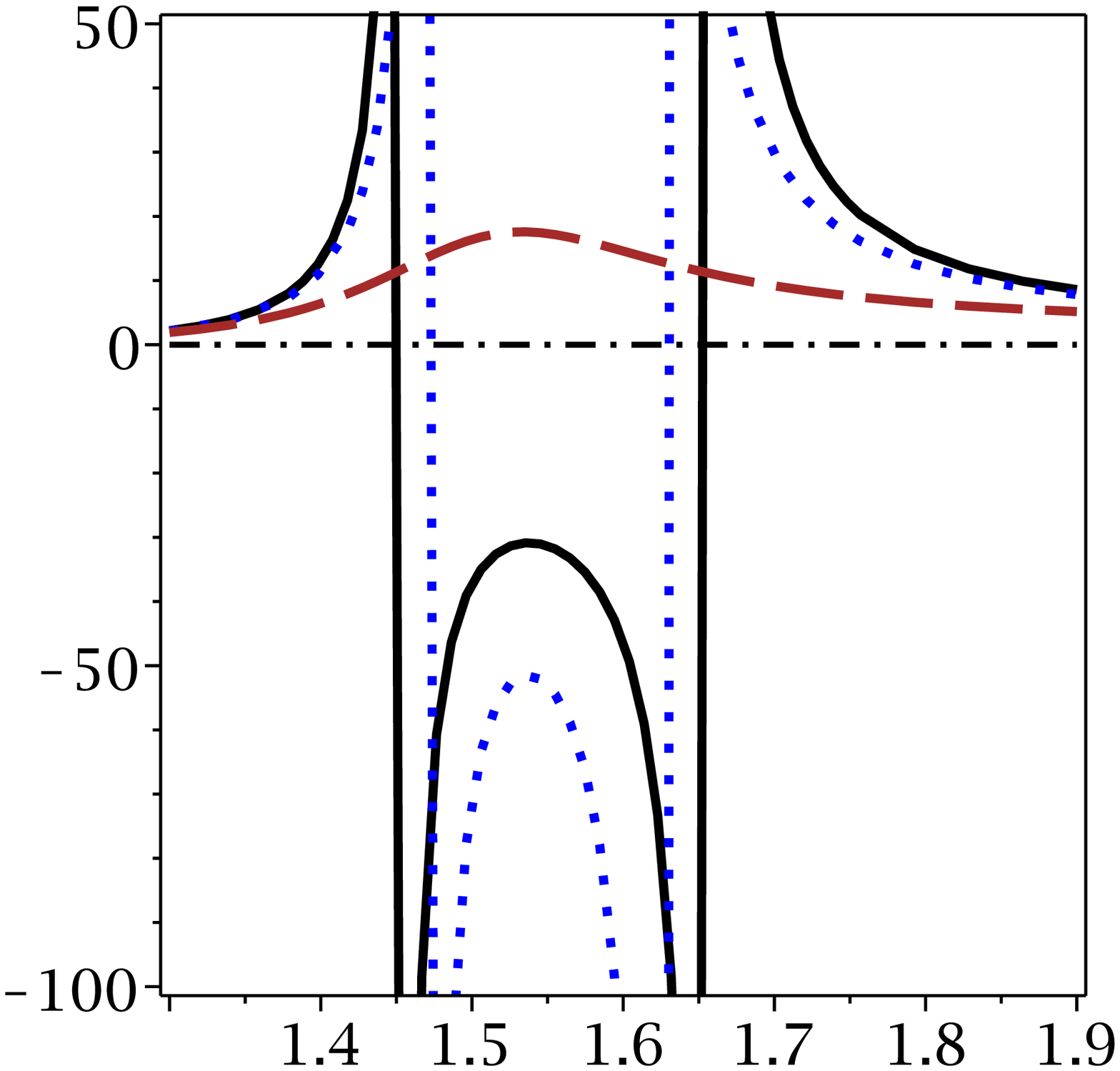}%
\end{array}
$%
\caption{For different scales: $C_{Q}$ versus $r_{+}$ for $\Lambda =-1$, $%
c=1 $, $d=5$, $q=1$ and $\protect\omega=0.01$ (continues line), $\protect%
\omega=0.1$ (dotted line) and $\protect\omega=1$ (dashed line). }
\label{Fig6}
\end{figure}


\begin{figure}[tbp]
$%
\begin{array}{ccc}
\epsfxsize=5cm \epsffile{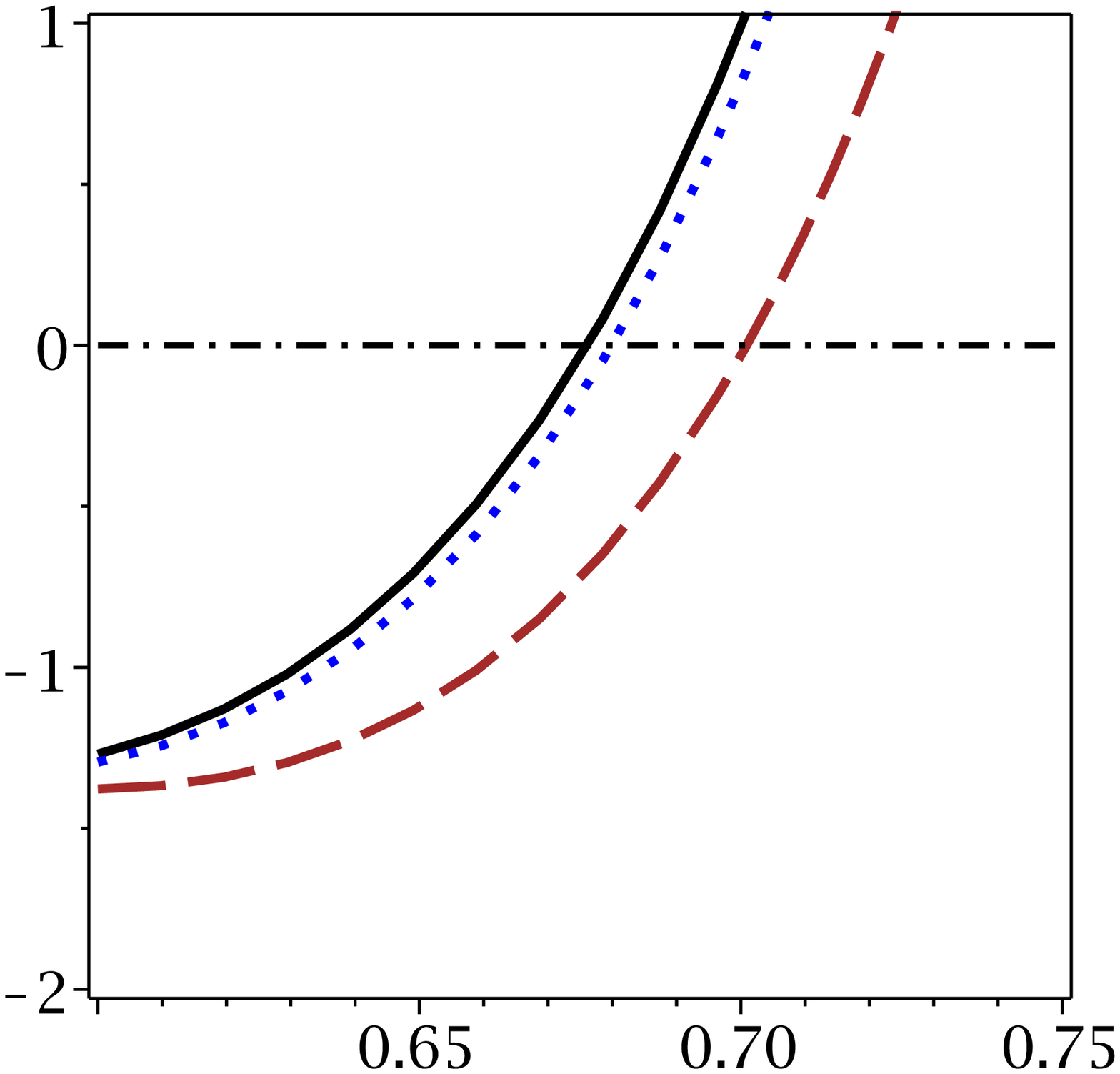} & \epsfxsize=5cm %
\epsffile{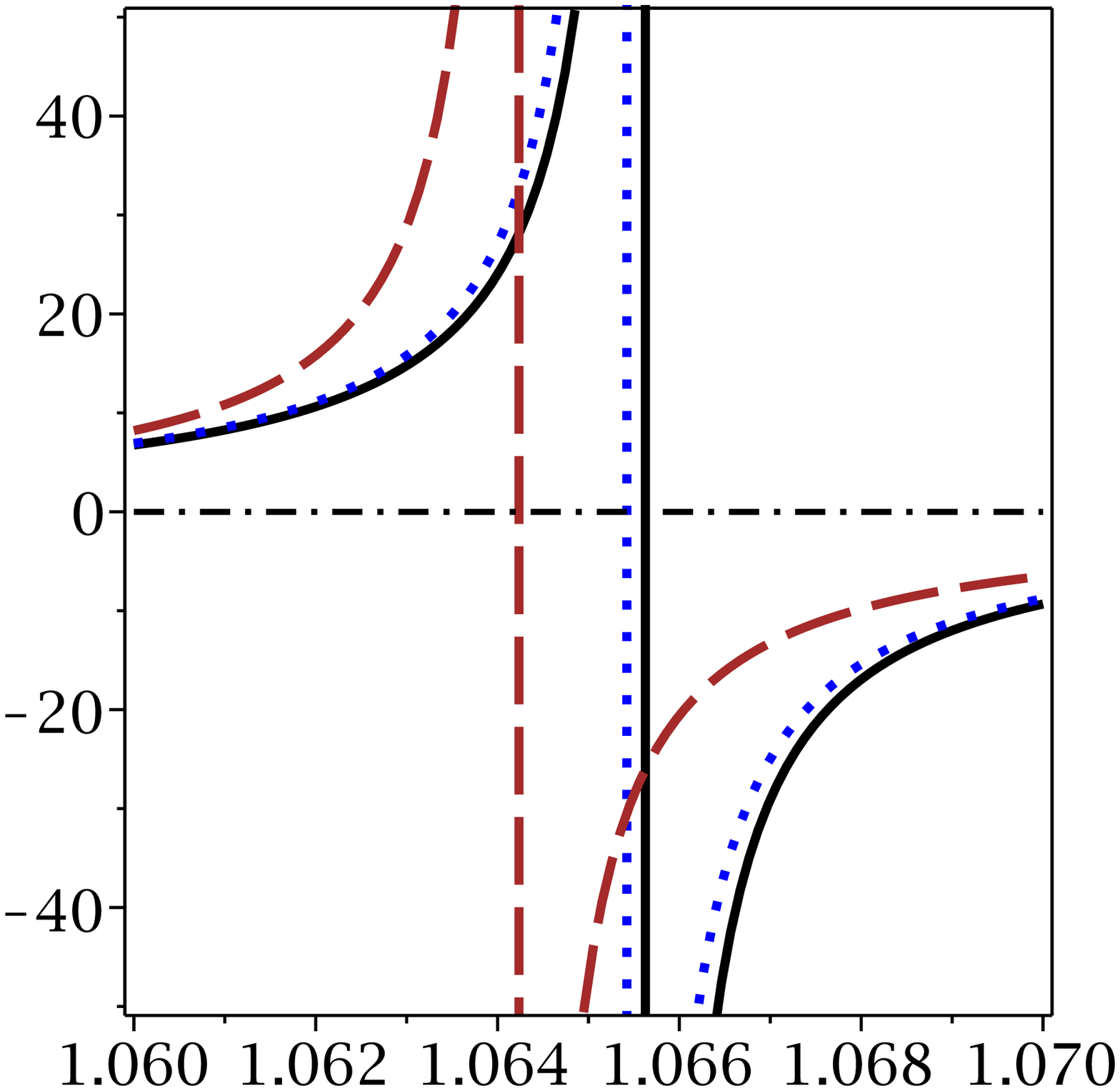} & \epsfxsize=5cm \epsffile{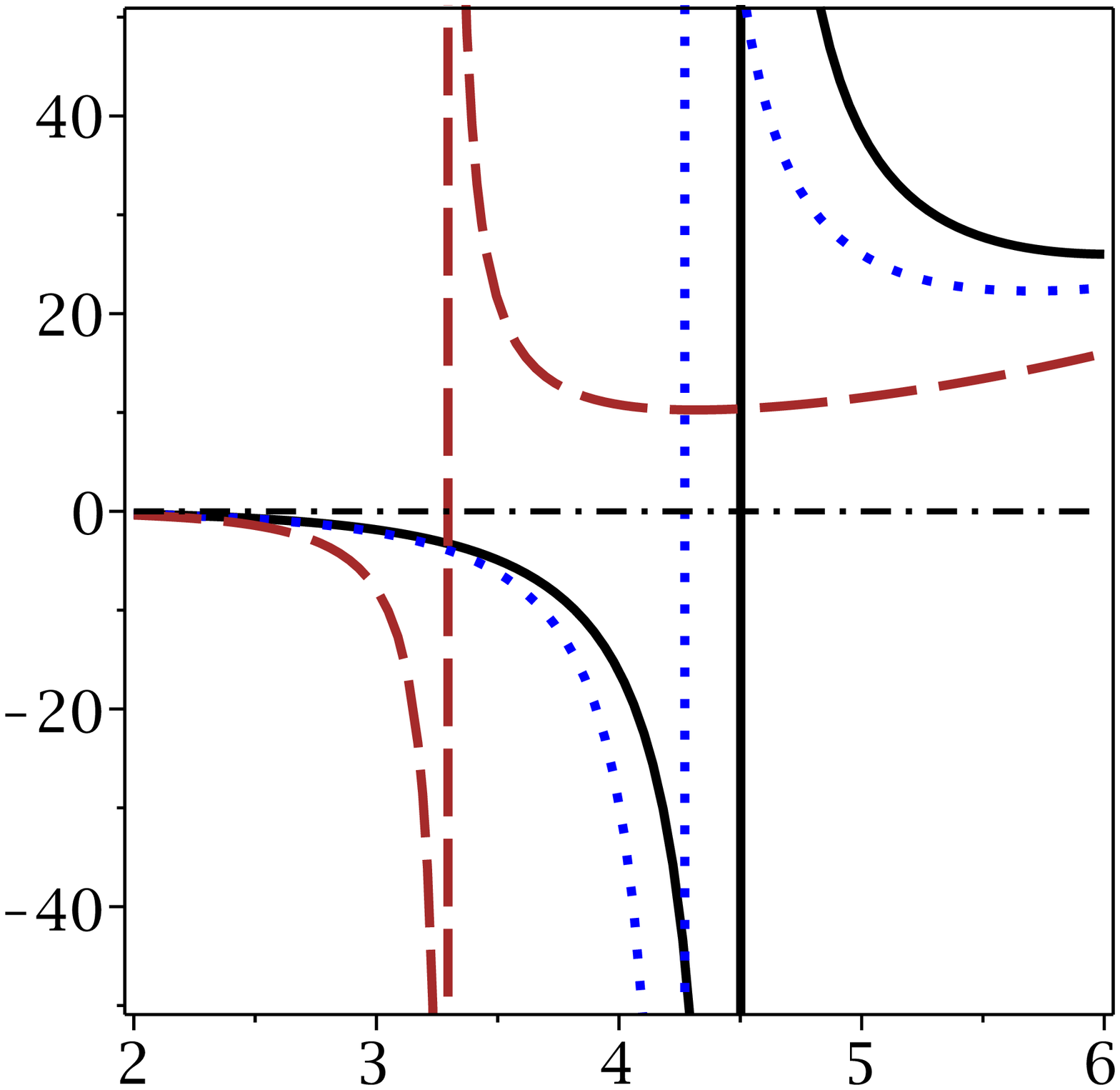}%
\end{array}
$%
\caption{For different scales: $C_{Q}$ versus $r_{+}$ for $\Lambda =-1$, $%
c=1 $, $d=6$, $q=1$ and $\protect\omega=0.01$ (continues line), $\protect%
\omega=0.1$ (dotted line) and $\protect\omega=1$ (dashed line). }
\label{Fig7}
\end{figure}


Regarding $C_{Q}$ versus entropy, numerical calculations and also figures
show that\ the heat capacity has a real positive root ($S_{0}$) for all
values of parameters ($q$, $\omega $, $\Lambda $, $c$). For the case of
divergence points of heat capacity, there are two cases. In first case and
for special choices of parameters, heat capacity does not diverge, while for
second case, one can set free parameters in such a way that $C_{Q}$ has two
real positive divergence points ($S_{1c}$ and $S_{2c}$).

In order to emphasize the effects of the Brans-Dicke gravity, we have
plotted Figs. \ref{Fig6} and \ref{Fig7} for variation of $\omega $. It is
evident that the variation of the Brans-Dicke parameter modifies the number
of the divergencies and roots and their corresponding places. These
modifications lead to changes in stability conditions (regions of the
stability) as well as phase transitions of the black holes. This shows that
in presence of modified gravity (Brans-Dicke gravity), the thermodynamical
structure of the black holes will be modified and acquires different
structure comparing to the absence of Brans-Dicke gravity.

According to the pioneering work of Davies \cite{Davies},
regarding phase transitions in black holes, the divergencies of
the heat capacity are second order phase transition. In his works,
he showed that these divergencies of the heat capacity have
characteristics of the second order phase transition. In addition,
the studies that are conducted in extended phase space proved that
divergencies of the heat capacity, and second order phase
transition in phase diagrams, coincide with each other
\cite{extended}. In other words, second order phase transition
points that are observed in Gibbs free energy versus temperature,
pressure versus horizon radius and temperature versus horizon
radius, are matched with divergencies of the heat capacity.
Therefore, the divergencies of the heat capacity are second order
phase transitions. For more clarifications, we will study
corresponding free energy versus horizon radius as well.

The free energy for these black holes is given by
\[
F=M-TS,
\]
in which by using Eqs. (\ref{m}), (\ref{s}) and (\ref{temp1}), one
can find free energy as
\begin{eqnarray}
F &=&\frac{\zeta \left[ n\left( n-2\right) +4\omega \left( n-1\right) \right]
}{4\pi j_{3}}\left\{ \frac{q^{2}\left[ n\left( \omega +\frac{9}{8}\right) -%
\frac{3}{2}\omega -\frac{9}{8}\right] }{\left[ \omega \left( n-1\right) +n%
\right] c^{j_{1}}r^{j_{3}}}+\frac{\left( n-2\right) c^{j_{1}}r^{j_{3}}}{4}%
\right.  \nonumber \\
&&  \nonumber \\
&&\left. -\frac{\Lambda \left( n-4\omega -9\right) \left[ n\left( \omega +%
\frac{5}{4}\right) -2\omega -\frac{9}{4}\right] c^{(n+1)j_{1}}r^{\zeta
\left( n-1\right) \left[ 3\left( 3n+1\right) +4n\omega \right] }}{2\left[
\omega \left( n-1\right) +n\right] \left[ n\left( 3+4\omega \right) +9\right]
}\right\} .
\end{eqnarray}

Now, by employing this relation and specific choices of different
parameters in plotted diagrams, we plot following figures
(\ref{Fig8}-\ref{Fig10}).

\begin{figure}[tbp]
$%
\begin{array}{cc}
\epsfxsize=6cm \epsffile{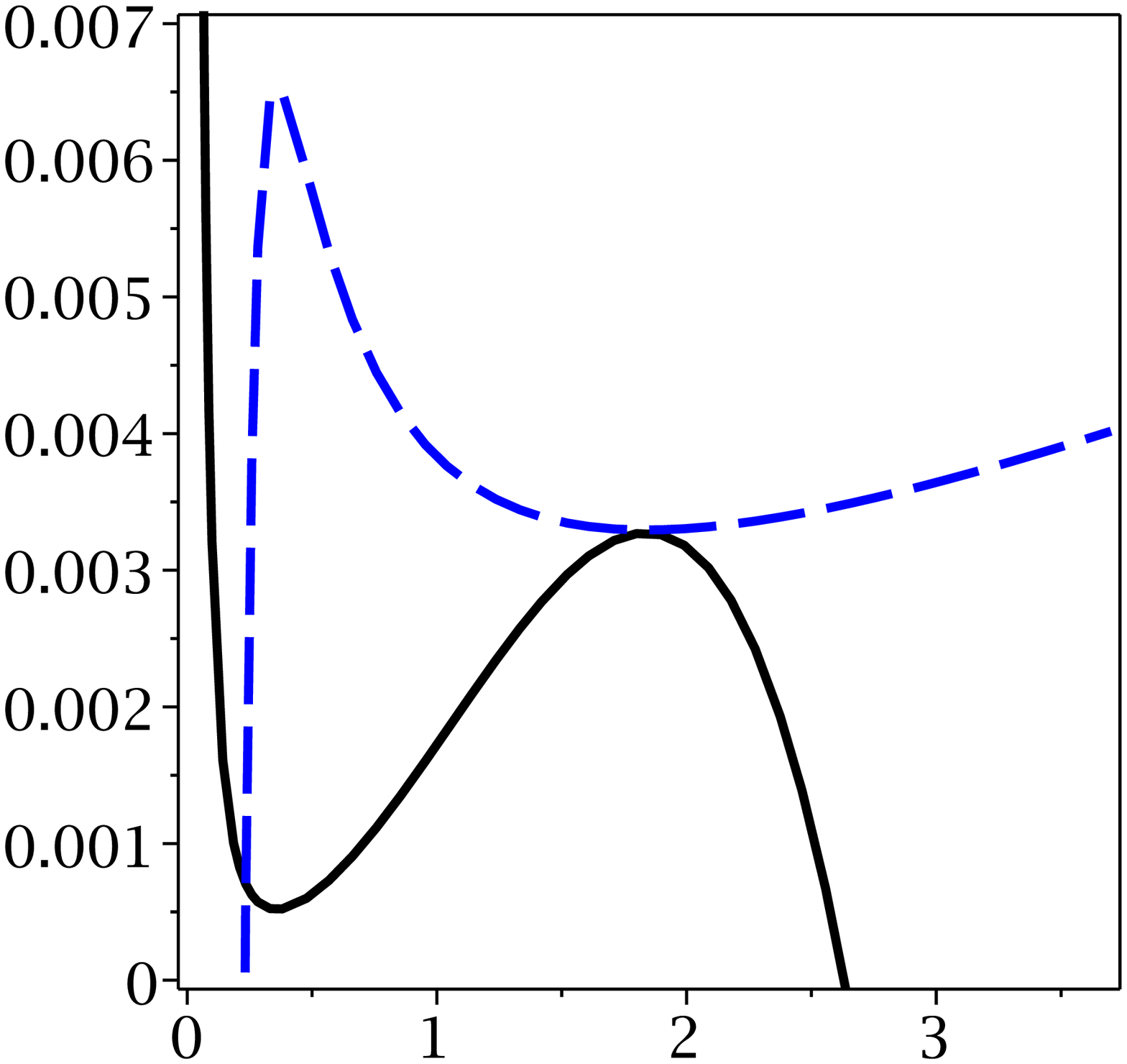} & \epsfxsize=6cm %
\epsffile{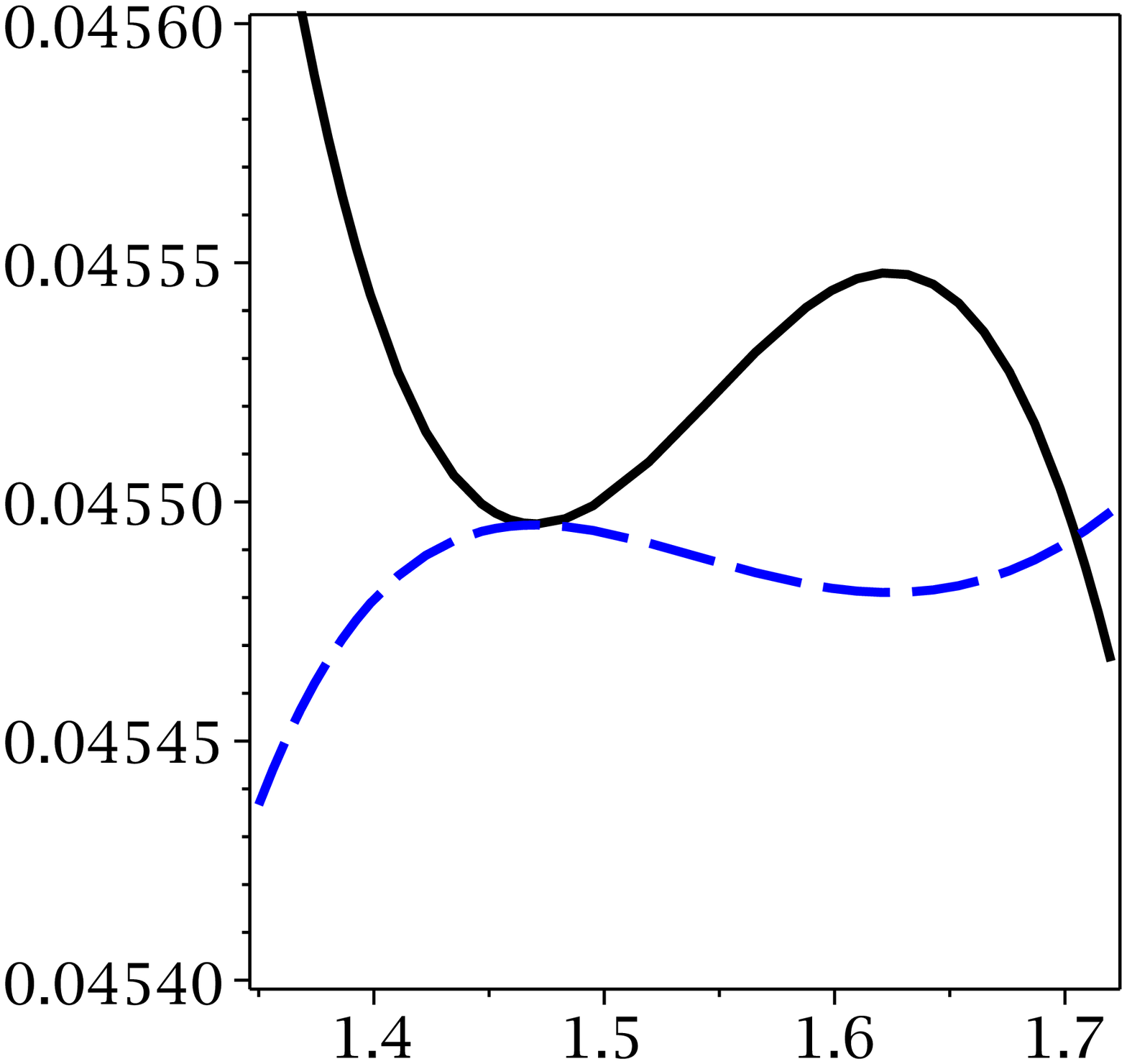}%
\end{array}
$%
\caption{For different scales: $F$ (continuous line) and $T$
(dashed line) versus $r_{+}$ for $\Lambda =-1$, $c=1$ and $d=5$.
(left diagram: $\protect\omega=1$ and $q=0.1$ \& right diagram:
$q=1$ and $\protect\omega=0.1$) } \label{Fig8}
\end{figure}

\begin{figure}[tbp]
$%
\begin{array}{ccc}
\epsfxsize=5cm \epsffile{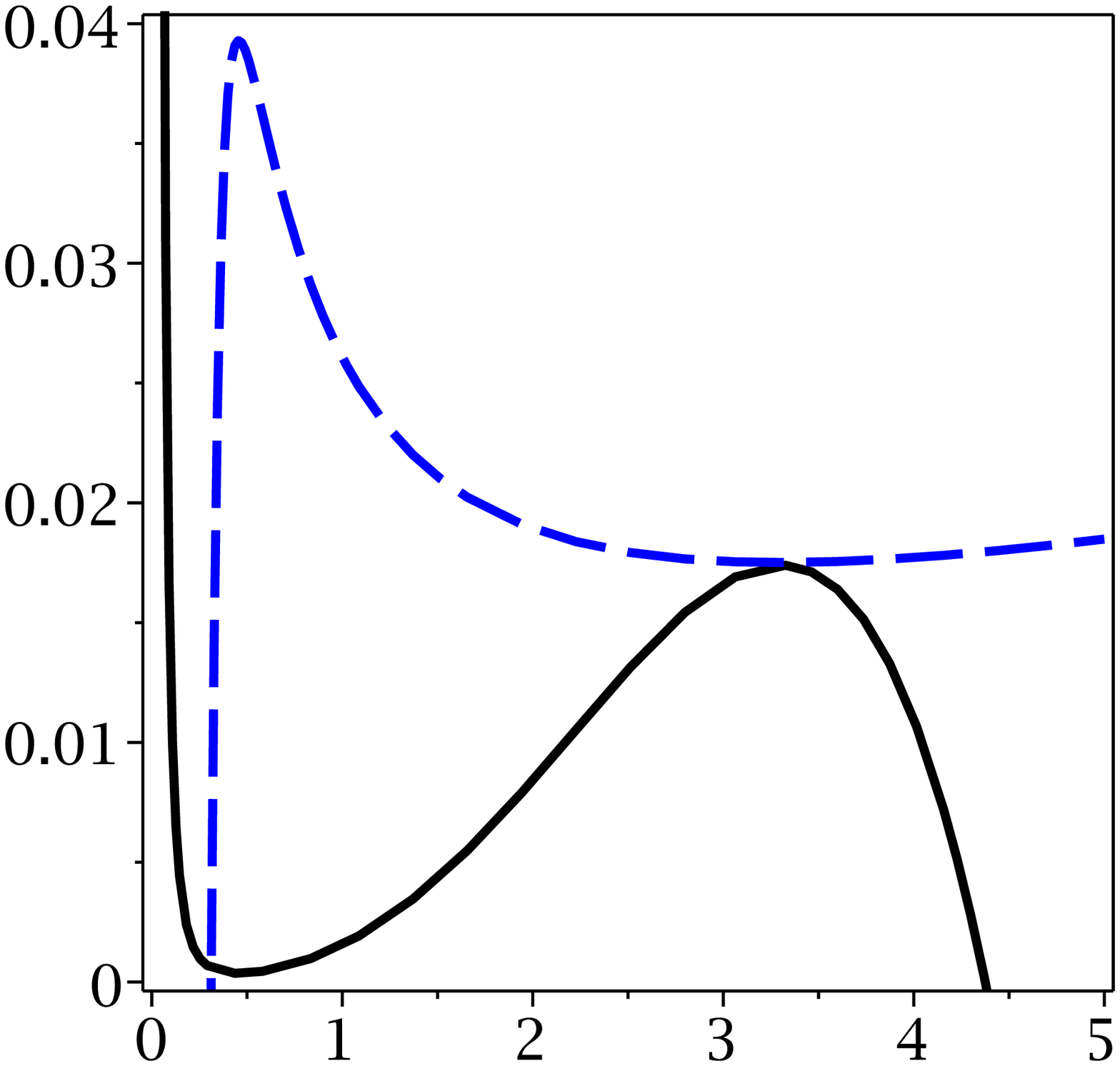} & \epsfxsize=5cm %
\epsffile{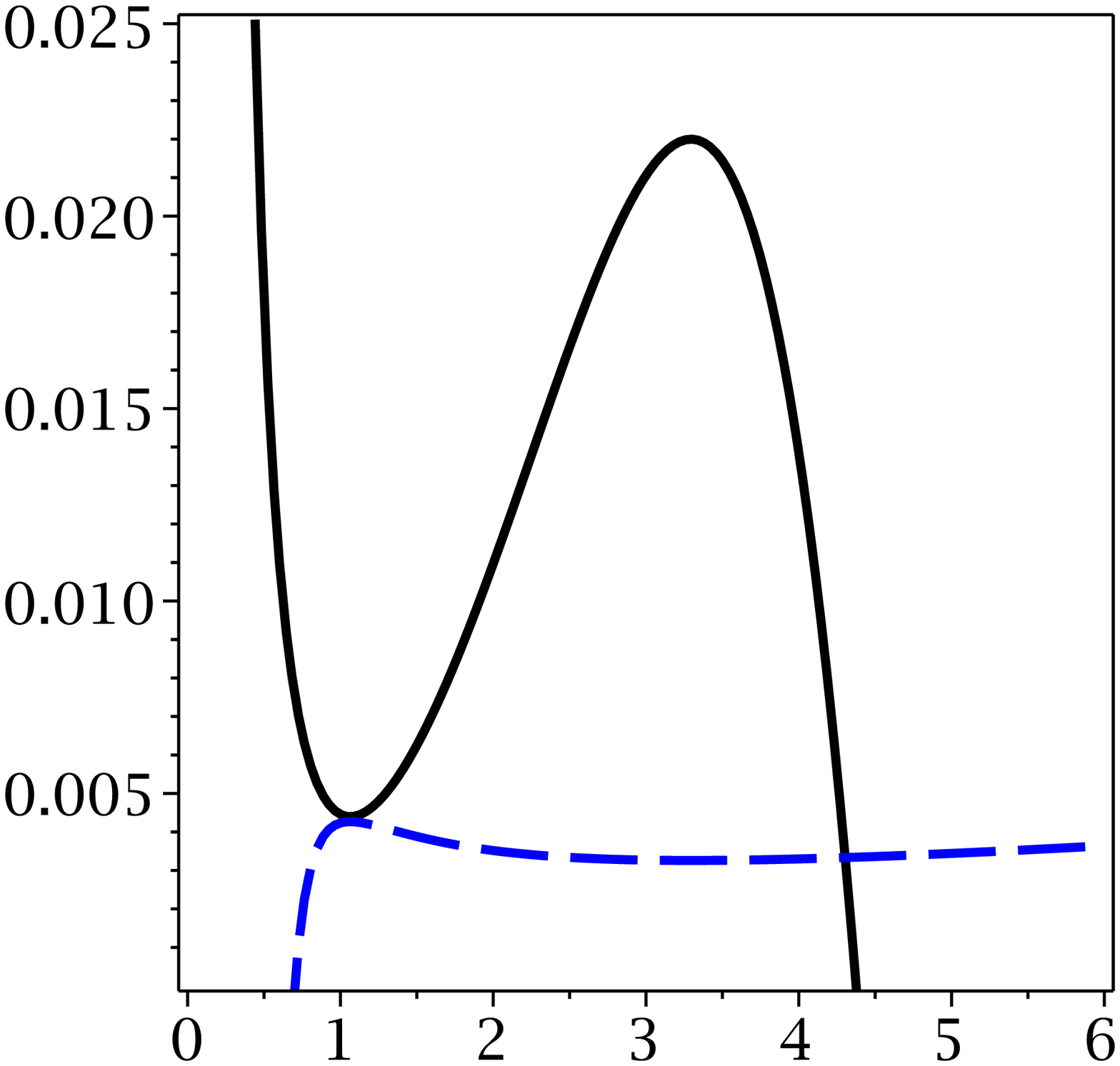} & \epsfxsize=5cm \epsffile{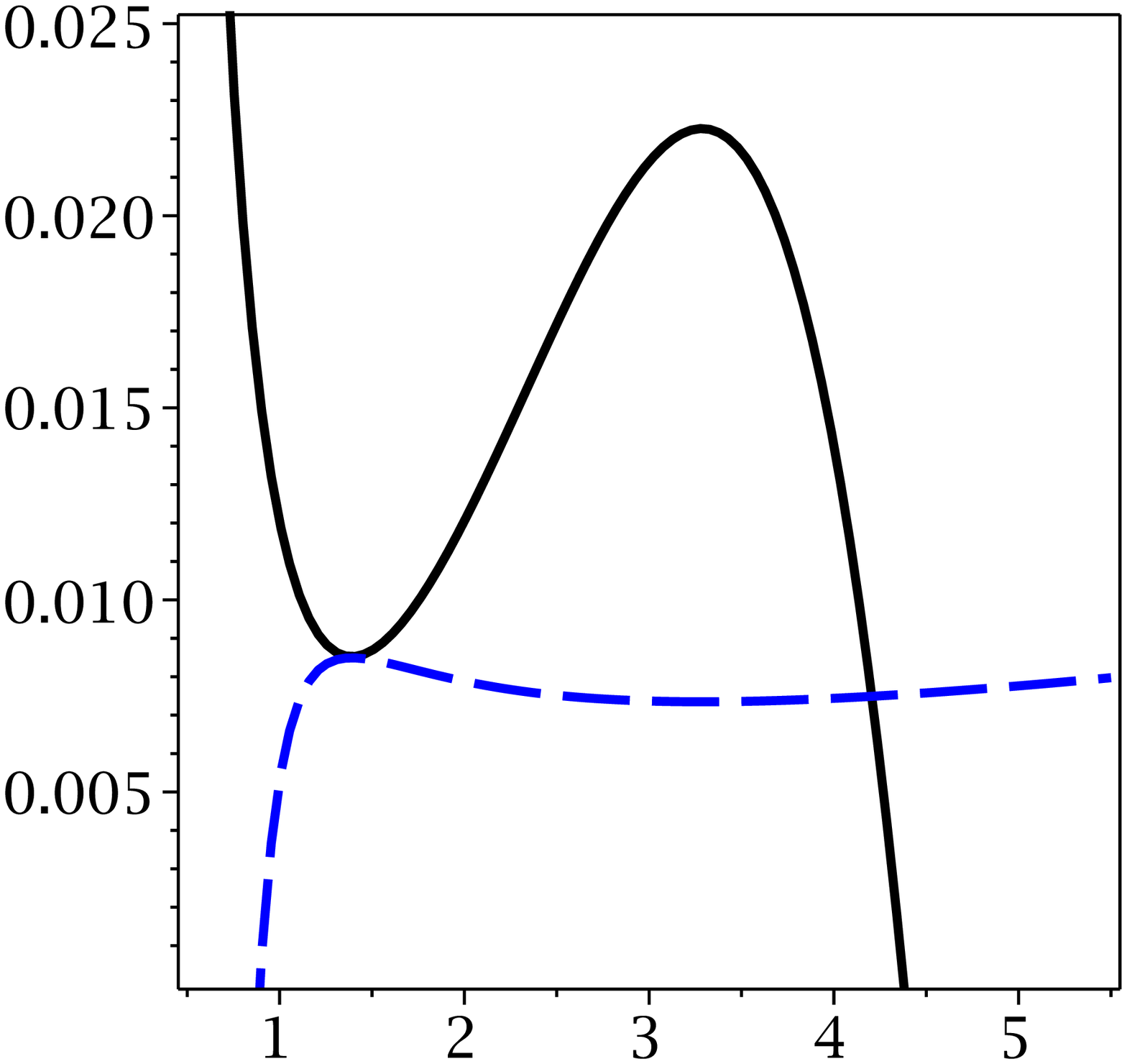}%
\end{array}
$%
\caption{For different scales: $F$ (continuous line) and $T$
(dashed line) versus $r_{+}$ for $\Lambda =-1$, $c=1$, $d=6$ and
$\protect\omega=1$. ( left diagram: $q=0.1$ \& middle diagram:
$q=1$ \& right diagram: $q=2$) } \label{Fig9}
\end{figure}

\begin{figure}[tbp]
$%
\begin{array}{cc}
\epsfxsize=6cm \epsffile{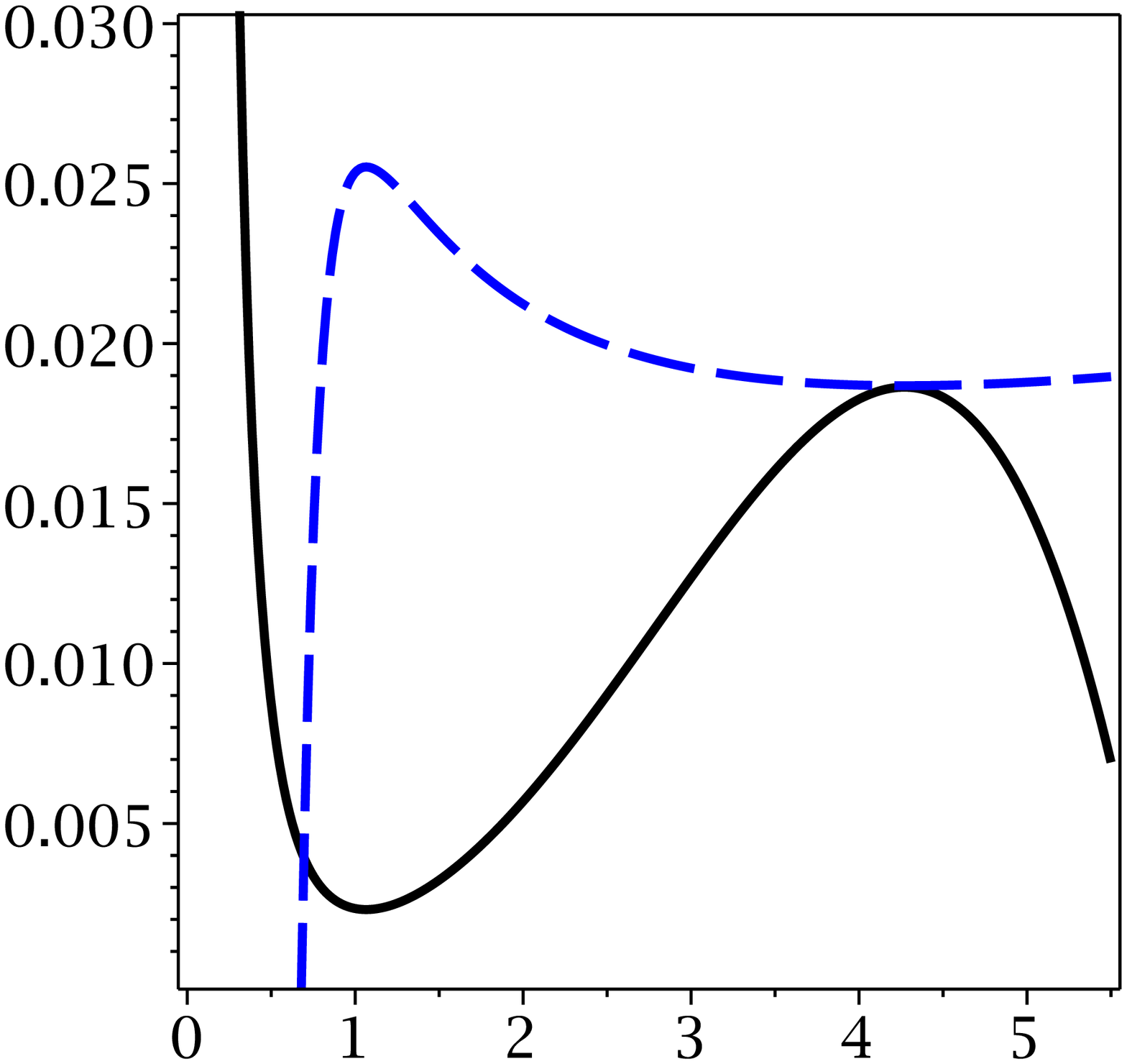} & \epsfxsize=6cm %
\epsffile{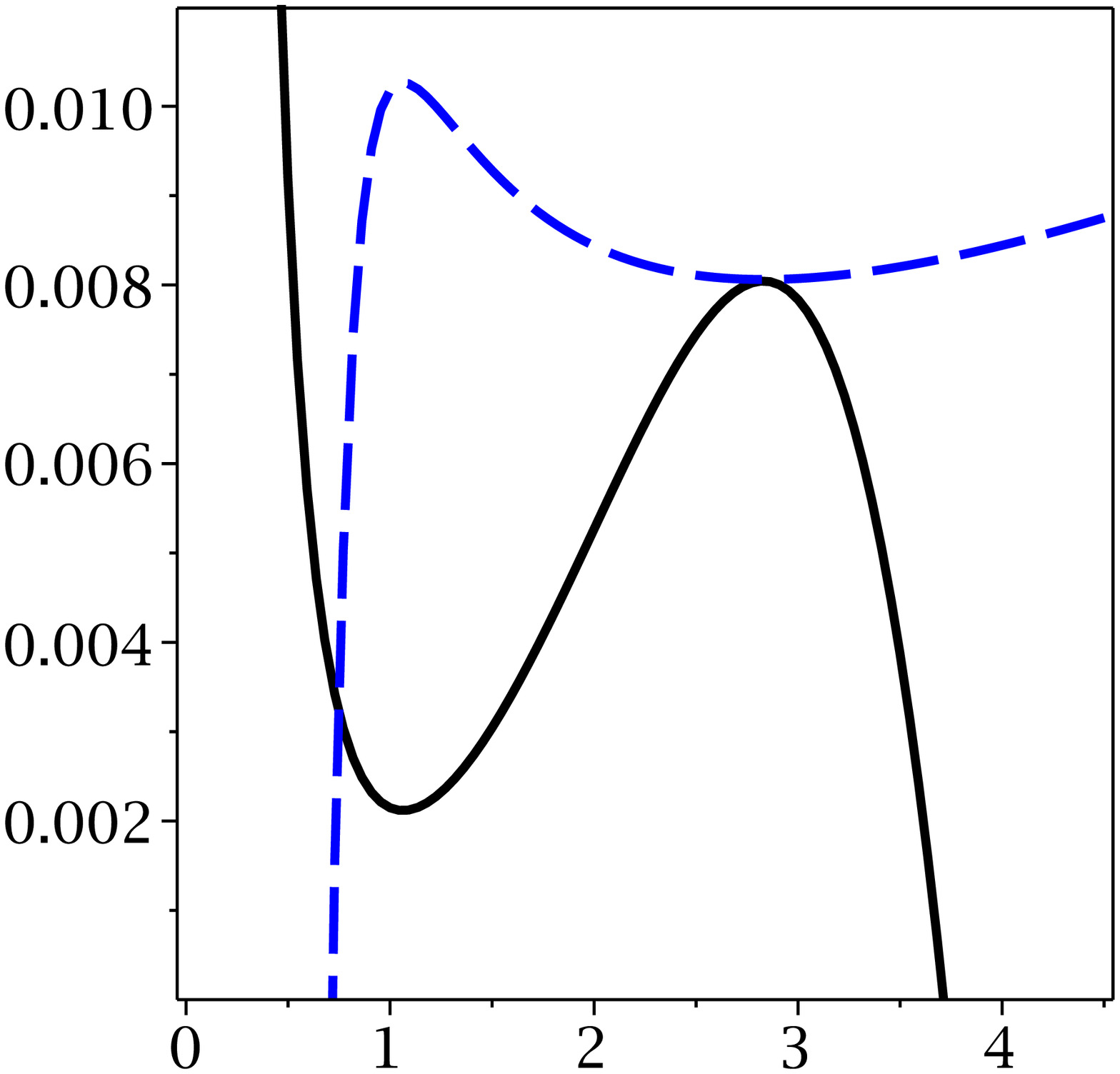}%
\end{array}
$%
\caption{For different scales: $F$ (continuous line) and $T$
(dashed line) versus $r_{+}$ for $\Lambda =-1$, $c=1$, $d=6$ and
$q=1$. ( left diagram: $\protect\omega=0.1$ \& right diagram:
$\protect\omega=3$) } \label{Fig10}
\end{figure}


Thermodynamically speaking, in free energy diagrams, the second
order phase transition takes place when the free energy acquires
an extremum. Here, we see for these specific choices of the
different parameters, free energy has two extrema. These extrema
are located exactly where temperature has extrema as well. The
existence of extrema in the temperature is observed as
divergencies in the heat capacity. Therefore, the extrema of free
energy, where second order phase transition takes place, coincide
with divergencies of the heat capacity. This leads to the
conclusion that divergencies that are observed in the heat
capacity are where black holes go under second order phase
transition.

Furthermore, we calculate critical horizon radius and temperature
by $\left( \frac{\partial T}{\partial r_{+}}\right)$ and present
the results in table II. In order to compare the results of heat
capacity with those of $T-r_{+}$ diagram and HPEM approach for
calculating the critical horizon radius, we combine all results in
table II. It is evident from obtained values for critical horizon
radius and their corresponding critical temperature that they
coincide with extrema in free energy and divergencies of the heat
capacity as well. This shows that phase transition points which
are second order ones are consistent for free energy, temperature
and heat capacity as well. Therefore, divergencies of the heat
capacity in fact are places in which second order phase
transitions take place.


\begin{center}
\begin{tabular}{c|c|c|c|c|c|c}
\hline\hline
$d$ & $q$ & $\omega $ & $%
\begin{array}{c}
T-r_{+}\text{ diagram:} \\
T_{c}\rightarrow
\end{array}%
$ & $%
\begin{array}{c}
T-r_{+}\text{ diagram:} \\
r_{c}\rightarrow
\end{array}%
$ & $%
\begin{array}{c}
F-r_{+}\text{ diagram:} \\
r_{c}\rightarrow
\end{array}%
$ & \multicolumn{1}{|c}{$%
\begin{array}{c}
C_{Q}-r_{+}\text{ diagram:} \\
r_{c}\rightarrow
\end{array}%
$} \\ \hline\hline
$5$ & $0.1$ & $1$ & $%
\begin{array}{c}
0.38486 \\
0.19380%
\end{array}%
$ & $%
\begin{array}{c}
0.35794 \\
1.83804%
\end{array}%
$ & $%
\begin{array}{c}
0.35794 \\
1.83804%
\end{array}%
$ & \multicolumn{1}{|c}{$%
\begin{array}{c}
0.35794 \\
1.83804%
\end{array}%
$} \\ \hline
$5$ & $1$ & $0.1$ & $%
\begin{array}{c}
0.19709 \\
0.19703%
\end{array}%
$ & $%
\begin{array}{c}
1.47014 \\
1.62412%
\end{array}%
$ & $%
\begin{array}{c}
1.47014 \\
1.62412%
\end{array}%
$ & \multicolumn{1}{|c}{$%
\begin{array}{c}
1.47014 \\
1.62412%
\end{array}%
$} \\ \hline
$6$ & $0.1$ & $1$ & $%
\begin{array}{c}
0.51557 \\
0.22972%
\end{array}%
$ & $%
\begin{array}{c}
0.45643 \\
3.29845%
\end{array}%
$ & $%
\begin{array}{c}
0.45643 \\
3.29845%
\end{array}%
$ & \multicolumn{1}{|c}{$%
\begin{array}{c}
0.45643 \\
3.29845%
\end{array}%
$} \\ \hline
$6$ & $1$ & $1$ & $%
\begin{array}{c}
0.30070 \\
0.22969%
\end{array}%
$ & $%
\begin{array}{c}
1.06423 \\
3.29541%
\end{array}%
$ & $%
\begin{array}{c}
1.06423 \\
3.29541%
\end{array}%
$ & \multicolumn{1}{|c}{$%
\begin{array}{c}
1.06423 \\
3.29541%
\end{array}%
$} \\ \hline
$6$ & $2$ & $1$ & $%
\begin{array}{c}
0.26534 \\
0.22963%
\end{array}%
$ & $%
\begin{array}{c}
1.39027 \\
3.28597%
\end{array}%
$ & $%
\begin{array}{c}
1.39027 \\
3.28597%
\end{array}%
$ & \multicolumn{1}{|c}{$%
\begin{array}{c}
1.39027 \\
3.28597%
\end{array}%
$} \\ \hline
$6$ & $1$ & $0.1$ & $%
\begin{array}{c}
0.34958 \\
0.25592%
\end{array}%
$ & $%
\begin{array}{c}
1.06542 \\
4.27029%
\end{array}%
$ & $%
\begin{array}{c}
1.06542 \\
4.27029%
\end{array}%
$ & \multicolumn{1}{|c}{$%
\begin{array}{c}
1.06542 \\
4.27029%
\end{array}%
$} \\ \hline
$6$ & $1$ & $3$ & $%
\begin{array}{c}
0.27082 \\
0.21270%
\end{array}%
$ & $%
\begin{array}{c}
1.06333 \\
2.83018%
\end{array}%
$ & $%
\begin{array}{c}
1.06333 \\
2.83018%
\end{array}%
$ & \multicolumn{1}{|c}{$%
\begin{array}{c}
1.06333 \\
2.83018%
\end{array}%
$} \\ \hline
\end{tabular}
\\[0pt]
{Table II: Critical horizon radius and temperature for $\Lambda =-1$ and $c=1
$.}
\end{center}


\section{Closing Remarks}

In this paper, we have studied the thermal stability and phase transitions
of charged BD black holes in context of canonical ensemble by calculating
the heat capacity. We showed that there is a lower bound for the horizon
radius of physical charged BD black holes. This restriction was originated
from the sign of the temperature of these black holes. We found that the
regions of the physical and non--physical black holes were functions of
electric charge and BD coupling constant.

Regarding the phase transition of the black holes, we found that black holes
in the context of BD enjoy the existence of second order phase transition.
In other words, the heat capacity of these black holes diverged in two
points and it had a real valued positive root. It was pointed out that the
existence of the divergence points and their places were functions of $q$
and $\omega $. It is worthwhile to mention that the effect of variation of
electric charge on larger divergence point was relatively so small. This
small effect indicates that the existence of the larger divergence point is
due to contribution of the BD gravity. It was also seen that dimensions
changed the existence of the divergence points of the heat capacity and also
the places of the root and divergencies of it. It is notable to mention that
these black holes have three characteristic points which are related to
positive temperature and thermal (in)stability of these black holes.

In the context of thermal stability, it was pointed out that there are four
regions with different conditions. These regions are specified by the root
and two divergencies of the heat capacity. In case of root of the heat
capacity, there was a limitation point between non-physical black holes and
physical ones. Between two divergencies, it was an unstable state and after
the larger divergency it acquired stable state. In other words, in smaller
divergency black hole may go under phase transition from unstable state with
larger horizon radius to smaller stable black hole, whereas, in case of
larger divergency system went under another phase transition and stabilized
with larger horizon radius.

Finally, we used the geometrical thermodynamics for studying phase
transitions of the system. It was shown that the divergencies of the
curvature scalar of the HPEM metric exactly coincide with both physical
limitation point and phase transitions of the heat capacity. In other words,
divergencies of the heat capacity and its root are matched with divergencies
of the Ricci scalar. It was also shown unlike RN-AdS black holes \cite{Niu}
(in the absence of dilaton field), employing Weinhold, Ruppeiner and Quevedo
metrics failed to provide effective results for BD black hole solutions. It
is notable that the behavior of the Ricci scalar around its divergence
points for root and phase transitions was different. It means that, there
were characteristic behaviors that enable one to recognize divergence point
of Ricci scalar related to root of the heat capacity from divergence points
of $\mathcal{R}$ related to divergencies of $C_{Q}$. Therefore, one is able
to point out that physical limitation point and phase transitions took place
in the divergencies of thermodynamical Ricci scalar of the HPEM metric with
different distinctive behaviors.

In addition, the free energy of these black holes has been
investigated as well. It was shown that free energy enjoys extrema
in its diagrams versus horizon radius. These extrema were located
exactly where temperature acquires extrema and heat capacity
diverges. It was pointed out that divergencies that are observed
in heat capacity are places where black holes go under second
order phase transition.

\section*{Acknowledgements}

We would like to thank the referees for their constructive comments. We also
thank the Shiraz University Research Council. This work has been supported
financially by the Research Institute for Astronomy and Astrophysics of
Maragha, Iran.

\end{document}